\documentclass[%
 preprint,
superscriptaddress,
 amsmath,amssymb,
 aps, physrev,
]{revtex4-2}

\usepackage{graphicx}
\usepackage{dcolumn}
\usepackage{bm}
\usepackage{amsmath}

\usepackage[english]{babel}

\usepackage{hyperref}
\hypersetup{pdfstartview={FitH},pdfpagemode={UseNone},
            colorlinks,linkcolor=blue, citecolor=blue, urlcolor=blue,
            bookmarksopen=true, pdfnewwindow=true}
\usepackage[all]{hypcap}
\usepackage[normalem]{ulem}

\begin{document}
\sloppy


\title{\textbf{Enhanced Tunable Photon Pair Generation from Nonlinear Metasurface with Guided-Mode Cavity} 
}%

\author{Tongmiao Fan}
 \email{Tongmiao.Fan@anu.edu.au}
\affiliation{
ARC Centre of Excellence for Transformative Meta-Optical Systems (TMOS), Department of Electronic Materials Engineering, Research School of Physics, Australian National University, Canberra, ACT 2600, Australia
}

\author{Jihua Zhang}
 
\affiliation{
 Songshan Lake Materials Laboratory, Dongguan, Guangdong 523808, P. R. China
}%
\affiliation{
ARC Centre of Excellence for Transformative Meta-Optical Systems (TMOS), Department of Electronic Materials Engineering, Research School of Physics, Australian National University, Canberra, ACT 2600, Australia
}

\author{Andrey A. Sukhorukov}
\affiliation{
ARC Centre of Excellence for Transformative Meta-Optical Systems (TMOS), Department of Electronic Materials Engineering, Research School of Physics, Australian National University, Canberra, ACT 2600, Australia
}

\date{\today}

\begin{abstract}
The ability to generate quantum entangled photon pairs through spontaneous
parametric down conversion (SPDC) is playing a pivotal role in many applications in quantum technologies, including quantum communications, quantum computation, and quantum imaging. Metasurfaces, two-dimensional arrays of nanostructures with subwavelength thickness, have recently shown extraordinary capability in manipulating classical and quantum states of light and realizing miniaturized SPDC sources.
Whereas previous studies have primarily focused on periodic metasurfaces, we uncover the potential of spatially modulated nanopatterns for further SPDC enhancement and state engineering. Specifically, we propose a nonlinear metasurface featuring a lateral guided-mode cavity formed by two distributed Bragg reflectors.
With the quasi-normal mode theory, we predict an SPDC rate up to 157~Hz/mW, which represents about 30 times rate enhancement when compared to the metasurface without a cavity.
Furthermore, the cavity supports a continuous set of high-Q resonances across a broad bandwidth, allowing the all-optical tuning of the photon-pair emission directions transverse to the cavity by controlling the pump wavelength, which can find applications in quantum imaging.
\end{abstract}

\maketitle


\section{Introduction}
Quantum photon pairs, which exhibit quantum entanglement, are fundamental to several emerging quantum technologies including quantum computing~\cite{Slussarenko:2019-41303:APPR}, quantum imaging~\cite{Clark:2021-60401:APL}, and quantum communications~\cite{Li2004}. 
One of the most prevalent techniques for generating quantum-entangled photon pairs with non-classical correlations is Spontaneous Parametric Down Conversion (SPDC), which utilizes nonlinear optics~\cite{Klyshko:1988:PhotonsNonlinear, Couteau:2018-291:CTMP}.
The SPDC process generates entangled photons by splitting a single photon into two lower-frequency entangled photons, termed the signal and idler. 
This phenomenon can occur due to second-order nonlinearity ($\chi^{(2)}$) in crystals. 
Despite the fact that bulky crystals can deliver relatively high efficiency, the requirement of extra temperature-control elements complicates the integration into compact optical devices~\cite{Wang:2020-273:NPHOT}. 
Furthermore, the bulky volume imposes strict longitudinal phase-matching condition to SPDC process, which limits the versatility of quantum states generation.

In recent years, researchers have turned their attention to developing new platforms, like nonlinear thin films, nanoresonators and quantum metasurfaces, for SPDC to overcome these previous limitations~\cite{Solntsev:2021-327:NPHOT, Kan:2023-2202759:ADOM, Ma:2024-2313589:ADM}. 
Multiple material systems have been explored 
such as metamaterials~\cite{Poddubny:2016-123901:PRL, Davoyan:2018-608:OPT}, 2D materials~\cite{Saleh:2018-3862:SRP, Guo:2023-53:NAT, Weissflog:2024-7600:NCOM} and liquid crystals~\cite{Sultanov:2024-294:NAT}. 
In particular, the platforms offering ultrasmall thickness 
can relax the strict longitudinal phase-matching condition to only a transverse phase-matching conservation~\cite{Okoth:2019-263602:PRL}, which enhances the flexibility of quantum photon generation and allows for strong spatial~\cite{Okoth:2020-11801:PRA}, time-frequency~\cite{Santiago-Cruz:2021-653:OL} and polarization~\cite{Sultanov:2022-3872:OL} entanglement.
However, thin films suffer from the low-efficiency due to the short interaction length.
Therefore, the optical nanoresonators and metasurfaces, where the nonlinear interactions can be strongly increased, have been actively explored for the enhancement of SPDC.
Nonlinear nanoresonators provide a versatile platform for photon-pair generation~\cite{Marino:2019-1416:OPT, Nikolaeva:2021-43703:PRA, Duong:2022-3696:OME, Zilli2023, Saerens:2023-3245:NANL, Olekhno:2024-245416:PRB, Weissflog:2024-11403:APPR}, allowing for the reconfigurable shaping of spatial emission patterns and engineering of the polarisation entanglement.

While single nanoresonators offer various benefits for SPDC, their practical application is still challenging due to the low photon-pair rates and a need for complex optical setups with highly focused pump beams.
This is where metasurfaces, which are arrays of nanoresonators, come into the picture~\cite{Kuznetsov:2024-816:ACSP, Schulz:2024-260701:APL}. Metasurfaces combine the advantages of nanoresonators with added benefits such as scalability to higher photon-pair rates and ease of integration.
Consequently, researchers are increasingly focusing on metasurfaces as a promising solution for the efficient and flexible generation of quantum photon pairs.
Indeed, nonlinear metasurfaces were shown in experiments to not only significantly enhance the photon-pair generation compared to thin-films~\cite{Santiago-Cruz:2021-4423:NANL}, but also facilitate the shaping of spatial~\cite{Zhang:2022-eabq4240:SCA} and spectral~\cite{Santiago-Cruz:2022-991:SCI} entanglement with polarizaion control~\cite{Ma:2023-8091:NANL, Jia:2025-eads3576:SCA} and tunable multi-directional emission~\cite{Weissflog:2024-3563:NANP}.


\begin{figure*}[hbt!]
    \centering
  \includegraphics[width=1\linewidth]{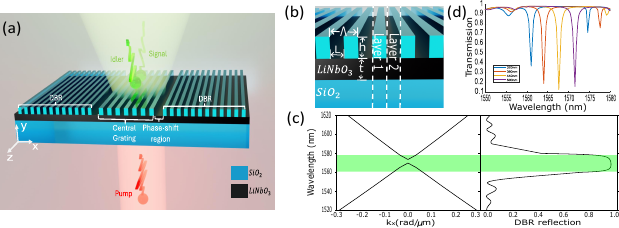}
  \caption{(a) Schematic view of the metasurface featuring DBRs on both sides. Gratings made up of silicon dioxide (SiO$_2$) are on top of a lithium niobate (LiNbO$_3$) thin film to enhance the electric field and form a guided-mode cavity. 
  Generated photons and pump photon are $z$-polarized.
  (b)~Schematic representation of two distinct layers composed of air, LiNbO$_3$, and SiO$_2$. Layer~1 includes an additional SiO$_2$ film on top of the LiNbO$_3$ film. 
  (c)~Wavelengths of guided mode resonance for the periodic central grating (left) and reflection of DBRs containing 150 periods (right). The green shaded area marks the wavelengths of high DBR reflectivity above 85\%.
  (d)~Simulated transmission through cavity metasurfaces versus various phase-shift region widths. The input beams from air-side of metasurfaces have a Gaussian profile along $x$ with a radius of 10 $\mu$m.}
  \label{design_of_parameters_new}
\end{figure*}

However, the maximum experimentally observed or theoretically predicted generation rate from the periodic metasurfaces
is still insufficient for many applications, being orders of magnitude below the rates from bulky crystal sources~\cite{Schulz:2024-260701:APL}. 
%
%
In this work, we suggest a new direction by introducing a transverse modulation of the metasurface, forming a cavity and supporting the light confinement for much stronger SPDC enhancement compared to a purely periodic metasurface.
We demonstrate our approach for a nonlocal metasurface featuring guided mode resonances~\cite{Zhang:2022-eabq4240:SCA} by truncating the meta-grating and sandwiching it between two distributed Bragg reflectors (DBRs) to form a transverse Fabry-Perot cavity, non-trivially extending a configuration that has been applied to enhance the classical second-harmonic generation (SHG)~\cite{Yuan:2021-153901:PRL}.
With this design, it becomes feasible to incorporate multiple cavities on a single chip, enabling a spatially multiplexed SPDC source—an arrangement unattainable with periodic metasurfaces.
We then apply quasi-normal mode (QNM) theory to rigorously predict an increase in generation rate by a factor of nearly~30, under identical pump conditions, while simultaneously reducing 
the metasurface dimensions.
%
Remarkably, while the cavity strongly confines light, it uniquely enables all-optical continuous tunability of photon spectra and emission angles in the transverse direction by simply controlling the pump wavelength. 
This feature represents a significant advantage over traditional nonlinear crystals, which typically require temperature tuning to have distinct emission spectra and angles, and also overcomes the constraints of quasi-bound state in the continuum (qBIC) modes in metasurfaces~\cite{Parry:2021-55001:ADP, Santiago-Cruz:2022-991:SCI, Mazzanti:2022-35006:NJP, Liu:2024-155424:PRB}.

\section{Quantum Metasurface with guided-mode cavity}

Incorporating a cavity into the metasurface design is expected to significantly enhance the electric field within it, leading to an increased photon pair generation rate. 
We consider a metasurface incorporating LiNbO$_3$, since this material is commonly used for the generation of quantum photon pairs due to its properties of large second-order susceptibility, low fluorescence, and high optical transmission in a broad wavelength range~\cite{Saravi:2021-2100789:ADOM}.
We start with a previously designed metasurface consisting of gratings made up of silicon dioxide (SiO$_2$) on top of a lithium niobate (LiNbO$_3$) thin film~\cite{Zhang:2022-eabq4240:SCA, Ma:2023-8091:NANL, Weissflog:2024-3563:NANP}, and then introduce 
distributed Bragg reflectors (DBRs)
acting as ``mirrors'' over a chosen wavelength range to tailor and enhance 
light-matter interactions in the central metasurface region as illustrated in Fig.~\ref{design_of_parameters_new}(a).
Utilizing DBRs as cavities in metasurface designs 
also benefits in terms of fabrication and integration~\cite{Lee:2020-4108:SENS}.

We target the photon-pair generation at the telecommunication wavelength range around 1570~nm. To facilitate the field enhancement through optical resonances at photon emission angles close to normal to the metasurface, 
the length of one period of central grating ($\Lambda_1$) is tuned to be 892.6~nm and the length of groove ($L_1$) to be 392.6~nm (filling ratio: 0.56) based on the eigenstudy in COMSOL Multiphysics.
The height of grating (h) is 200~nm, and the thickness of LN film (t) is 304~nm.
This configuration is specifically designed to facilitate the excitation of guided mode resonances propagating transversely through the structure, while emitting photons in the normal direction.
The number of periods for the central grating is designed as 20 to not only support the resonant mode but also to reasonably reduce the region of photon generation and emission.
For the emitted photons (~1570 nm), the operation remains within the subwavelength regime in both air and the substrate. Although the period exceeds the wavelength of the pump (~785 nm), this does not disqualify our design as a metasurface, as there are no engineered resonances at the pump wavelength.

The performance of cavity metasurface also depends on the reflection wavelength of the DBRs. 
In order to achieve reflection around the 1570~nm wavelength, the period of DBR ($\Lambda_2$) is engineered to be 445.88~nm, filling ratio to be 0.49. 
We design the period of DBR by $p = {\lambda}/({4 n_{eff1}})+{\lambda}/({4 n_{eff2}})$~\cite{Sheppard1995}, where $n_{eff1}$ and $n_{eff2}$ are effective refractive indices of two constructing layers as depicted in Fig.~\ref{design_of_parameters_new}(b), whose effective refractive indices are 1.7785 and 1.743, respectively. 
The resonant wavelengths of the guided modes supported by the periodic central grating, as depicted in Fig.~\ref{design_of_parameters_new}(c, left), confirm that the guided mode 
cannot propagate within the DBRs within the shaded region.
The reflection spectrum of DBRs shown in Fig.~\ref{design_of_parameters_new}(c, right) is calculated in COMSOL based on the boundary mode analysis through the excitation of the guided mode.
The reflectivity depends on the DBR size, and we choose the number of periods for DBR on one side as 150 to maximize the performance of guided-mode cavity while still keeping miniaturized dimension of the overall structure. 

Another important parameter that determines the resonant wavelength is the distance between the DBR and central grating, referred as a phase-shift region indicated in Fig.~\ref{design_of_parameters_new}(a). 
An optimal phase-shift of 440~nm was selected for having a cavity resonance around the telecommunication wavelength of 1570~nm based on the transmission simulation versus different phase-shifts as shown in Fig.~\ref{design_of_parameters_new}(d).
This brings the total width of the designed metasurface with a cavity to approximately 150~$\mu$m.
Due to the broad overall transverse dimension compared to a single unit cell, all COMSOL simulations were conducted using a 2D model to optimize the computation time. We performed sets of such simulations for different values of the out-of-plane wave number $\boldsymbol{k}_{z}$, which then allows us to obtain a complete solution in 3D space by obtaining the parameters of quasi-normal modes and then performing the spatial Fourier sums of 2D simulations.

\begin{figure*}[hbt!]
    \centering
  \includegraphics[width=1\linewidth]{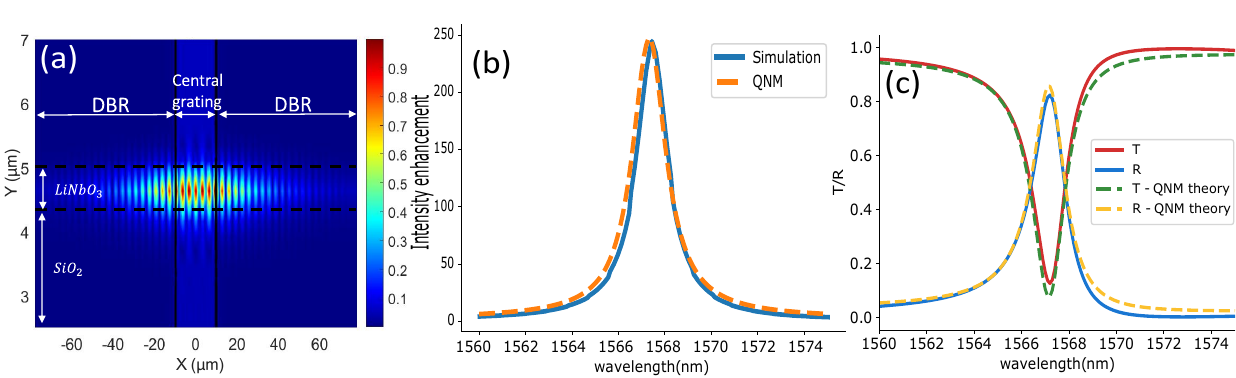}
  \caption{(a)~Normalized absolute values of the electric field for the dominant cavity mode at $1567.3$~nm, having the highest intensity within the centre of nonlinear LiNbO$_3$ layer. 
  (b) Enhancement of electric field amplitude in the nonlinear LiNbO$_3$ film relative to the peak amplitude of an incident beam calculated using QNM theory (dashed line) and direct COMSOL simulation(solid line). (c)~Transmission ($T$) and reflection ($R$) from the cavity.
  (b,c)~Calculations are based on QNM theory (dashed line) and direct COMSOL simulations (solid line), confirming high accuracy of QNM approach. The input beams have a Gaussian profile along $x$ with a radius of 10 $\mu$m, incident from the air-side on the metasurface.
  For all the plots, $\boldsymbol{k}_{z} = 0$. }
  \label{linear_QNM}
\end{figure*}

\section{Enhanced Tunable SPDC modelled by quasi-normal mode theory}

\subsection{Basic principles of quasi-normal modes}

Modes play a vital role in quantum physics, as they form the basis for the most comprehensive description of a system’s evolution through the superposition principle. 
Conservative systems are characterized by straightforward boundary conditions, and the operators that describe them are Hermitian or self-adjoint.
The eigenstates of such systems, often referred to as normal modes, constitute a complete set, and their corresponding eigenvalues, such as frequencies or energies, are real.
However, realistic open systems like metasurfaces are inherently non-Hermitian or non-conservative, and they are described by non-self-adjoint operators or non-Hermitian matrices. 
This results in modes not forming a complete set, leading to the designation of ``quasi''. Such eigenmodes of non-Hermitian open systems as conventionally referred to as quasi-normal modes~\cite{Lai:1990-5187:PRA, Lalanne:2018-1700113:LPR, Sauvan:2013-237401:PRL, Gigli:2020-1197:ACSP}, or QNMs for short.

For our metasurface, we seek electromagnetic QNMs in the form $\left[\widetilde{\mathbf{E}}_{m,\boldsymbol{k}_{z}}(\mathbf{r}), \widetilde{\mathbf{H}}_{m,\boldsymbol{k}_{z}}(\mathbf{r})\right]$ 
at a specific out-of-plane wave number $\boldsymbol{k}_{z}$. They are source-free solutions of Maxwell's equations for the permittivity distribution of the metasurface $\varepsilon(\mathbf{r}, \omega)$, at the eigenfrequency $\widetilde{\omega}_{m,\boldsymbol{k}_{z}}$~\cite{Lalanne:2018-1700113:LPR}:
%
\begin{equation}
\begin{split}
\label{maxwell}
\left[\begin{array}{cc}
0 & i \varepsilon^{-1}\left(r, \tilde{\omega}\right) \nabla \times \\
-i \mu^{-1}\left(r, \tilde{\omega}\right) \nabla \times & 0
\end{array}\right]
& \left[\begin{array}{c}
\tilde{\boldsymbol{E}}_{m,\boldsymbol{k}_{z}}(r) \\
\tilde{\boldsymbol{H}}_{m,\boldsymbol{k}_{z}}(r)
\end{array}\right]\\
=  \tilde{\omega}_{m,\boldsymbol{k}_{z}}
& \left[\begin{array}{c}
\tilde{\boldsymbol{E}}_{m,\boldsymbol{k}_{z}}(r) \\
\tilde{\boldsymbol{H}}_{m,\boldsymbol{k}_{z}}(r)
\end{array}\right] ,
\end{split}
\end{equation}
%
with outgoing-wave boundary conditions.

The Maxwell equation can be directly solved through eigenstudy in COMSOL, where the out-of-plane wave number $\boldsymbol{k}_{z}$ can be directly specified. The open character of the eigenproblem results in an unusual yet critical feature of QNMs, being that the field distributions $\left[\tilde{\boldsymbol{E}}_{m,\boldsymbol{k}_{z}}(r), \tilde{\boldsymbol{H}}_{m,\boldsymbol{k}_{z}}(r)\right]$ should diverge for $|y| \rightarrow \infty$ or $|x| \rightarrow \infty$. In order to normalize the QNM modes, we apply perfectly matched layer (PML) surrounding the overall metasurface in COMSOL. 
We can then normalize the QNM modes based on the following equation in a 2D model~\cite{Sauvan:2013-237401:PRL}:
\begin{align}
\iint  {\left[\tilde{\boldsymbol{E}}_{m,\boldsymbol{k}_{z}} \cdot \frac{\partial \omega \varepsilon}{\partial \omega} \tilde{\boldsymbol{E}}_{m,\boldsymbol{k}_{z}}\right.}
 \left.-\tilde{\boldsymbol{H}}_{m,\boldsymbol{k}_{z}} \cdot \frac{\partial \omega \mu}{\partial \omega} \tilde{\boldsymbol{H}}_{m,\boldsymbol{k}_{z}}\right] d\boldsymbol{x}d\boldsymbol{y}=1
 \label{normal-pml}
\end{align}
where $\epsilon$ and $\mu$ are the material permittivity and permeability, respectively. The materials are set to be non-dispersive in our simulations.

\subsection{Linear properties from QNM theory}

Through direct simulation and normalization in COMSOL Multiphysics, we deliberately select one QNM of the engineered metasurface with a strong field localization inside the cavity. While the mode exists for a range of $\boldsymbol{k}_{z}$, we shown in Fig.~\ref{linear_QNM}(a) a representative example at $\boldsymbol{k}_{z}=0$. The mode Q-factor is around 939. The electric field is confined to the centre of the nonlinear LiNbO$_3$ layer, which perfectly satisfies our design objective. The confinement of the field within the nonlinear layer of the metasurface cavity is a crucial factor in our design, as it indicates the potential of the cavity to effectively enhance photon-pair generation through SPDC, as we confirm in the following. 

The QNM theory significantly reduces computational time and offers greater physical insight into the scattering problem compared to direct numerical simulations. By selecting the appropriate scatterer and knowing the incident and background fields, the scattered field can be determined through numerical calculations. For our metasurface, background is made up of SiO$_2$ substrate and air, both the LiNbO$_3$ layer and gratings are selected as scatterers. The scattering problem can be solved using the following equations~\cite{Lalanne:2018-1700113:LPR}:
\begin{equation}
\left\{
\begin{aligned}
E_{t o t}&=E_b+E_s=E_b+\sum \alpha_{m,\boldsymbol{k}_{z}} \tilde{\boldsymbol{E}}_{m,\boldsymbol{k}_{z}} , \\
\alpha_{m,\boldsymbol{k}_{z}}(\omega)&=\iint_{S_{\mathrm{res}}}\left. \frac{\omega \, \Delta \varepsilon(\mathbf{r}, \omega)}{\tilde{\omega}_{m,\boldsymbol{k}_{z}}-\omega}\right. \mathbf{E}_b \cdot \tilde{\boldsymbol{E}}_{m,\boldsymbol{k}_{z}} d\boldsymbol{x}d\boldsymbol{y} .
\end{aligned}
\right.
\label{recons_eq}
\end{equation}
Here, $\Delta \varepsilon(\mathbf{r}, \omega)$ is the $\varepsilon$ difference between the background and scatter (LiNbO$_3$ and gratings), $E_{s}$ is the scattered field, $E_b$ is the background field in SiO$_2$ and air, and $E_{tot}$ is the total field, which is the sum of the scattered and background fields.

We now calculate the intensity enhancement in the cavity, defined as ${|E_{LN}|}^2/{|E_{inc}|}^2$, where $|E_{LN}|$ is the maximal electric field amplitude in the LiNbO$_3$ film, and $|E_{inc}|$ is the electric field amplitude of incident beam with a Gaussian profile in the $x$ direction.
We obtain an excellent agreement of the results from direct numerical simulations with COMSOL and quasi-normal theory based on Eq.~(\ref{recons_eq})
as depicted in Fig.~\ref{linear_QNM}(b).
We indeed observe that the cavity supports high intensity enhancement, reaching $250$ at resonance. This enhancement is particularly important for the SPDC process, as a stronger field can boost the nonlinear process in lithium niobate layer and the generation of photon pairs, making the overall system more efficient. We also check that the transmission and reflection are reconstructed by quasi-normal theory with high accuracy compared to numerical simulations as shown in Fig.~\ref{linear_QNM}(c).
This confirms that at the wavelengths of interest, response of the metasurface is dominated by the cavity mode, which we will apply below in the modeling of the nonlinear SPDC process.

\subsection{Photon pair generation rate from QNM theory}

The QNM theory already finds extensive applications in the modelling of nonlinear processes, including the second harmonic generation (SHG)~\cite{Gigli:2020-1197:ACSP} and SPDC~\cite{Weissflog:2024-11403:APPR}.
By employing the QNMs, we are able to accurately describe the nonlinear behavior of open systems, which is essential for understanding and optimizing the performance of nonlinear optical devices.

The SPDC generation rate can be rigorously modelled through the two-photon transition amplitude $\tilde{T}_{i s}$, which has the meaning of the complex wave function fully defining the pure two-photon state~\cite{Poddubny:2016-123901:PRL, Marino:2019-1416:OPT, Weissflog:2024-11403:APPR}:
\begin{align}
    \frac{\mathrm{d} N_{\text {pair }}}{d t d \omega_s }&=\frac{c^2 n_i n_s}{8 \omega_i \omega_s \pi^3}\int dS_i \int dS_s\left|\tilde{T}_{i s}\left(r_i, \omega_i, \sigma_i ; r_s, \omega_s, \sigma_s\right)\right|^2 ,
\end{align}
where the two-photon wave function $\tilde{T}$ is defined as
\begin{align}
    \tilde{T}_{i s}\left(r_s, r_i, \sigma_s, \sigma_i\right)&=\int \mathrm{d}^3 r_0 G_{\sigma_i \alpha}\left(r_i, r_0, \omega_i\right) G_{\sigma_s \beta}\left(r_s, r_0, \omega_s\right) \Gamma_{\alpha \beta}\left(r_0\right) .
    \label{rate-real-space}
\end{align}
Here $r_{i}$, $r_s$ are the positions where the signal and idler photons are detected, and $\omega_s$, $\omega_i$ are the corresponding frequencies of the signal and idler. $G_{\sigma_s \beta}$ is an electromagnetic Green function, where $\sigma$ denotes the polarization states of a photon. In the QNM basis, the Green function can be expressed as~\cite{Lalanne:2018-1700113:LPR}:
\begin{align}
    G\left(r, r_0, \omega\right)
    =-\sum_{m=1}^{\infty} 4\pi \epsilon_0\omega
    \frac{\tilde{E}_{m,\boldsymbol{k}_{z}}(r) \otimes \tilde{E}_{m,\boldsymbol{k}_{z}}\left(r_0\right)}{\left(\omega-\tilde{\omega}_{m,\boldsymbol{k}_{z}}\right)} .
    \label{green-non-periodc}
\end{align}
Here $\tilde{E}_{m,\boldsymbol{k}_{z}}$ is the normalized electric field of QNM with a specific transverse wavevector component $\boldsymbol{k}_{z}$, $\tilde{\omega}_{m,\boldsymbol{k}_{z}}$ is the corresponding frequency, and  $\Gamma_{\alpha \beta}$ is the generation matrix:
\begin{align}
    \Gamma_{\alpha \beta}=\chi_{\alpha \beta \gamma}^{(2)}\left(r_0 ; \omega_s, \omega_i ; \omega_p\right) E_{p, \gamma}(r_0) e^{-i \omega_p t} ,
\end{align}
where $E_{p, \gamma}$ is the electric field in the metasurface at the pump frequency $\omega_p$.

In this research, we employed Eq. (\ref{rate-real-space}) in conjunction with the corresponding Green function within the Quasi-Normal Mode (QNM) basis, as illustrated in Eq. (\ref{green-non-periodc}). As we stated, the behavior of the engineered metasurface cavity is dominated by the cavity mode as shown in Fig.~\ref{linear_QNM}(a), and we focus on this dominant mode to elucidate the underlying physics and evaluate the photon-pair emission properties.
Since \( |T(\boldsymbol{k}_x, \boldsymbol{k}_z, \omega_s)|^2 \) represents the photon-pair generation rate, the field enhancement within the nonlinear LiNbO\(_3\) film is expected to result in an increased SPDC rate. This field enhancement can be characterized by the normalized QNM field \( \tilde{E} \) in Fig. \ref{linear_QNM}(a).


\section{Results}

\begin{figure*}[hbt!]
    \centering
  \includegraphics[width=1\linewidth]{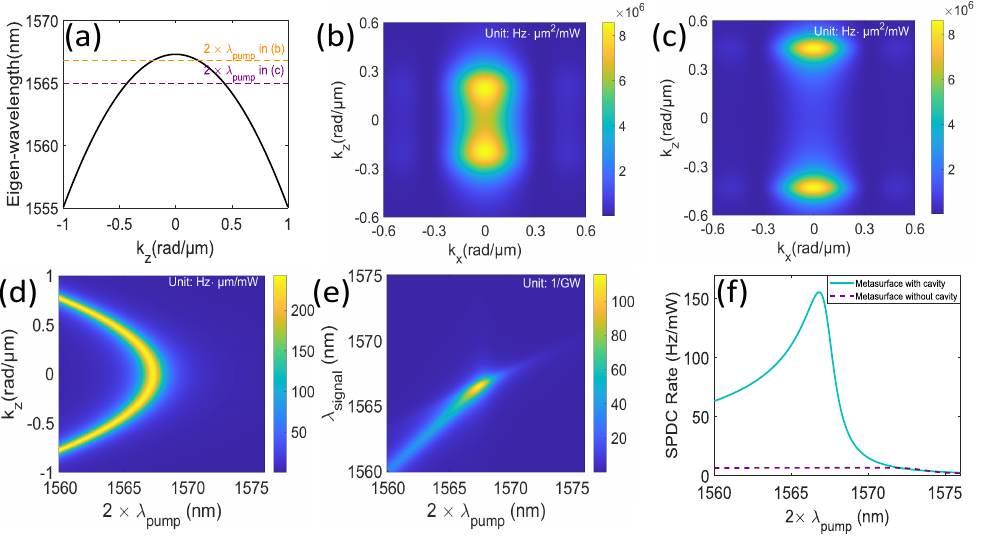}
  \caption{(a) Dispersion of the cavity eigen-wavelength vs. the transverse wavenumbers ($\boldsymbol{k}_z$). Over the shown wavelength range, all modes have a high quality factor of $Q \simeq 950$. Horizontal dashed line marks the wavelengths $2\times \lambda_{pump}=1566.8$~nm and $1565.2$~nm. 
  (b,c)~Spatial emission patterns of photons, integrated over the frequencies, corresponding to the different pump wavelengths as indicated by labels and dashed lines in~(a).
  (d)~Photon generation rate at different emission angles along the $z$ direction, integrated over $\boldsymbol{k}_x$ and photon frequencies, vs. the pump wavelength.
  (e)~Spectra of the signal photons for different pump frequencies, integrated over the transverse wavenumbers. (f)~Predicted total photon pair generation rate vs. the photon wavelength controlled by the pump. Our cavity structure (solid line) increases the rate by $\sim$30 times compared to an ideally periodic metasurface (dashed line).
  For all the plots in (b-f), the pump beam is incident from the substrate-side on the metasurface and has a Gaussian profile with 6-$\mu$m radius along $x$ and $z$.}
  \label{nonlinear_QNM}
\end{figure*}

The designed cavity supports a family of localized modes for varying transverse wavenumbers $\boldsymbol{k}_z$ along the $z$ direction. 
We find that within the range of $\boldsymbol{k}_z$ from $\Gamma$ point to $\pm$1 $rad/\mu m$ ($\pm$ 30 degrees), the resonant wavelengths can be shifted by over 10~nm as shown in the dispersion curve in Fig.~\ref{nonlinear_QNM}(a). 
This indicates that our structure supports a continuous set of resonances, allowing for an emission of quantum photon pairs at various angles and wavelengths when the energy conservation 
\begin{align}
&\omega_{pump}=\omega_{signal}+\omega_{idler}
\end{align}
and the transverse phase-matching condition 
\begin{align}
&\boldsymbol{k}_{p,z}=\boldsymbol{k}_{s,z}+\boldsymbol{k}_{i,z}
\end{align}
are met. Here $\omega$ represents the frequency of the pump $(p)$, signal $(s)$, and idler $(i)$ photons, while $\boldsymbol{k}_{z}$ represents their wavevector components along the $z$ direction.
For example, for two normally incident pump beams of different wavelengths, as indicated by dashed lines in Fig.~\ref{nonlinear_QNM}(a), the generated photon-pairs should have near-degenerate wavelengths at twice the pump wavelength. Then, their transverse wavenumbers and corresponding emission angles would also be different according to the dispersion dependence. Furthermore, there appears a possibility to tailor the biphoton spectra by spatially shaping the pump beam in $\boldsymbol{k}$-space
based on the general SPDC principles~\cite{Shukhin:2024-10158:OE}. 
Importantly, the quality factors (Q-factors) of all modes at different $\boldsymbol{k}_z$ values remain around 950, demonstrating that a significant enhancement in the efficiency of photon pair generation can potentially be achieved across a broad range. 



\begin{figure}[htbp]
\centering
\includegraphics[width=0.7\columnwidth]{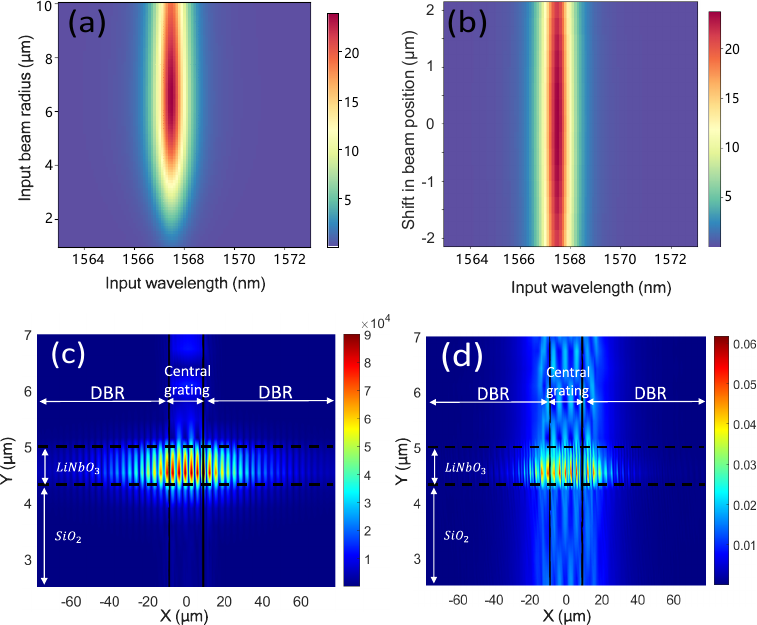}
\caption{(a,b) The second harmonic generation (SHG) rates (unit: $PW/m$) from the designed cavity metasurface for an input beam having a Gaussian profile along $x$ and incident from the air-side with different input wavelengths and (a)~beams with varying radii and positioned at the centre of the cavity or (b)~beams with a fixed 6-$\mu$m radius with a shift in $x$ direction from the cavity centre, as indicated by labels. (c,d)~Absolute value of the electric field of the fundamental field with a 6~$\mu$m input beam radius at 1567.2 nm wavelength and the generated second-harmonic field, respectively (Unit: $V/m$).}
\label{SHG}
\end{figure}

%
For efficient nonlinear photon conversion in the cavity, it is also important to determine the optimal beam radius. For this purpose, we simulated second harmonic generation (SHG), which can be regarded as the reverse processes of the frequency-degenerate SPDC~\cite{Parry:2021-55001:ADP}, from the cavity metasurface using nonlinear simulation in COMSOL, as depicted in Fig.~\ref{SHG}(a).
Ideally, the beam radius should be sufficiently large to excite the localized mode. However, if the radius is too large, outer parts of the beam would illuminate the DBR region, resulting in inefficient power utilization. 
The results presented in Fig.~\ref{SHG}(a) indicate that the fundamental beam radius of 6~$\mu$m, incident on the metasurface from the air side, achieves optimal SHG performance.
We also conduct simulations on SHG with a 6-$\mu$m-radius beam shifted from the center of the metasurface cavity along the $x$-axis. 
As shown in Fig.~\ref{SHG}(b), the rate remains high even with a 2-$\mu$m shift.
In accordance with the quantum-classical correspondence principle~\cite{Parry:2021-55001:ADP}, the rate of SHG serves as a reliable indicator of the SPDC rate, which thereby should be resilient to changes in beam radius or position.
The absolute values of the fundamental field for a 6-$\mu$m input beam at the cavity resonance wavelength and the generated second-harmonic are presented in Figs.~\ref{SHG}(c) and (d), respectively.
The confinement of electric field in the LiNbO$_3$ layer confirms the excitation of the target cavity mode.

We now proceed to the investigation of photon-pair generation through SPDC using the quasi-normal theory formulated in the previous section. We consider in the following the pump beams having a Gaussian profile with a radius of 6~$\mu$m in both the transverse directions $x$ and $z$, and the electric fields induced by the pump inside the nonlinear metasurface cavity are directly simulated in COMSOL Multiphysics.
By taking a Fourier transform of the two-photon wave function $\tilde{T}$ in real space in Eq.~(\ref{rate-real-space}), we obtain the wave function in $\boldsymbol{k}$-space, $T\left(\boldsymbol{k}_x,\boldsymbol{k}_z,\omega_s\right)=\int d x \exp \left(-i k_x x\right) T(x, k_z, \omega_s)$~\cite{Poddubny:2016-123901:PRL}. The signal-idler generation efficiency in all different regimes is then proportional to $|T\left(\boldsymbol{k}_x,\boldsymbol{k}_z,\omega_s\right)|^2$. 

First, we examine the spatial emission patterns of photon-pairs by integrating $|T\left(\boldsymbol{k}_x,\boldsymbol{k}_z,\omega_s\right)|^2$ over the frequency spectra with 50~nm bandwidth around the cavity resonance. As an example, we consider two normally incident pump beams with different wavelengths of $2\times \lambda_{pump}=1565.2$~nm and $1566.8$~nm, as indicated by dashed lines in Fig.~\ref{nonlinear_QNM}(a). 
As depicted in Figs.~\ref{nonlinear_QNM}(b,c), the predicted photon-pair emission patterns exhibit distinct maxima along the $z$-direction. The $\boldsymbol{k}_z$
position of the maximum emission rate aligns with the intersection of the corresponding dashed lines with the dispersion curve shown in Fig.~\ref{nonlinear_QNM}(a), in agreement with the transverse phase-matching principle. 
The width of bright spots in emission patterns along the $\boldsymbol{k}_x$ direction is inversely proportional to the width of the central grating in the cavity metasurface, while the widths along the $\boldsymbol{k}_z$ direction are inversely proportional to the Gaussian pump width in the $z$-direction. 
Most significantly, these results confirm the possibility of all-optical tuning of the transverse emission directions simply by controlling the pump wavelength.

We further analyze the properties of the emitted photons by calculating their emission pattern 
along $\boldsymbol{k}_z$ for pump wavelengths over a broad range, as depicted in Fig.~\ref{nonlinear_QNM}(d), by integrating $|T\left(\boldsymbol{k}_x,\boldsymbol{k}_z,\omega_s\right)|^2$ over $\boldsymbol{k}_x$ and $\omega_s$. 
The alignment of the emission maxima in Fig.~\ref{nonlinear_QNM}(d) with the dispersion curve in Fig.~\ref{nonlinear_QNM}(a) confirms that our structure can precisely control the angle and wavelength of emitted photons by adjusting the pump beam wavelength, which is an important feature in applications of quantum imaging~\cite{Pittman:1995-3429:PRA, Moreau:2019-367:NRP, Ma:2025-2:ELI}.

We now present in Fig.~\ref{nonlinear_QNM}(e) the spectra of signal photons at various pump frequencies, by integrating $|T\left(\boldsymbol{k}_x,\boldsymbol{k}_z,\omega_s\right)|^2$ over $\boldsymbol{k}_x$ and $\boldsymbol{k}_z$.
It is consistent with the findings in the periodic meta-grating metasurface~\cite{Zhang:2022-eabq4240:SCA} that the degenerate case, where signal and idler are at the same wavelength, exhibits a higher generation rate in agreement with the transverse phase-matching condition. However, in the case of our designed metasurface, the resonance within the cavity leads to a narrower bandwidth of the generated photons of about 1~nm, compared to the bandwidth of 2~nm in the ideally periodic meta-grating metasurface~\cite{Zhang:2022-eabq4240:SCA}. Additionally, a further enhancement in the telecommunication band around 1567~nm highlights the effectiveness of our design. The generated quantum photon pairs may be suitable for many quantum application like quantum key distribution (QKD)~\cite{Griffiths:2023-44040:PRAP}.

Finally, to assess the overall generation rate, we integrate $|T\left(\boldsymbol{k}_x,\boldsymbol{k}_z,\omega_s\right)|^2$ over the $\boldsymbol{k}$-space range of $( \pm 1 ) rad/\mu m$, corresponding to a detection angle of approximately 30~degrees, and photon frequencies $\omega_s$ over a 50~nm bandwidth. The calculated total SPDC rate is shown in Fig.~\ref{nonlinear_QNM}(f). 
For comparison, we also estimate the total SPDC rate from an ideally periodic metasurface without DBR regions, using the coupled-mode theory (CMT) formulated previously~\cite{Zhang:2022-eabq4240:SCA}. We find that, as shown in Fig.~\ref{nonlinear_QNM}(f), our cavity structure increases the rate by over one order of magnitude compared to a simple periodic metasurface.
This significant enhancement underscores the advantages of our metasurface design in terms of both tunability and rate.


To better contextualize the performance of our proposed cavity, we compare it against state-of-the-art metasurface platforms developed for enhancing SPDC. 
An experimental realization of a GaAs metasurface supporting a single qBIC resonance has demonstrated an enhancement of approximately $10^3$ over a wider 100-nm photon bandwidth~\cite{Santiago-Cruz:2022-991:SCI}. 
Several theoretical designs have also shown substantial enhancement in simulations. One approach utilizes a dual-resonance configuration in an AlGaAs metasurface, where two quasi-BIC modes are engineered to coincide with the signal and idler frequencies, yielding an enhancement factor of approximately $10^3$ over a narrow 6-nm photon bandwidth~\cite{Liu:2024-155424:PRB}. Another proposal employs a dual-BIC strategy in an Al$_{0.18}$Ga$_{0.82}$As metasurface, predicting a comparable enhancement ($\sim10^3$) over a broader estimated 50-nm photon bandwidth~\cite{Parry:2021-55001:ADP}. The enhancement in Ref.~\cite{Parry:2021-55001:ADP} is estimated based on the common assumptions that (i)~the spectral response generated by the thin film exhibits a spectrally uniform profile, and (ii)~the SPDC rate from a thin film coincides with the off-resonant SPDC rate of the metasurface.
Hybrid plasmonic metasurfaces offer another strategy for enhancing SPDC by leveraging localized field enhancement. In one such design, an enhancement factor of up to $\sim500$ was estimated over a 40-nm bandwidth, based on the reported field enhancement in numerical simulations~\cite{Jin:2021-19903:NASC}.
In comparison, our architecture achieves an SPDC enhancement of $\sim10^4$ over a 50-nm photon bandwidth, which is an order of magnitude higher than in previously reported metasurface platforms. This enhancement is realized using a simple heterostructure without relying on qBIC engineering or plasmonic confinement. The resulting structure is not only highly effective, but also fabrication-friendly and scalable, providing a promising pathway for integrated quantum photonics.

\section{Conclusion}

In conclusion, we have designed a metasurface featuring a guided-mode cavity facilitating electric field concentration in a quadratically nonlinear lithium niobate film at the resonant wavelengths in the telecommunication band around the $1550$~nm wavelength.
Remarkably, our newly engineered metasurface can resonantly enhance the photon pair generation rate by $\sim$30 times relative to the previously analysed simple periodic metasurface, while the tunability of emission angles via pump wavelength is preserved.
This level of enhancement surpasses that achieved by most designs based on qBIC resonances~\cite{Liu:2024-155424:PRB,Parry:2021-55001:ADP,Santiago-Cruz:2022-991:SCI}.
This significant enhancement is maintained over a 10-nm bandwidth of pump wavelengths, offering versatile all-optical reconfigurability for a range of applications. For instance, the ability to tune the emission angles of spatially entangled photons transversely to the grating direction at continuously varying pump wavelengths from the metasurface can be leveraged in quantum imaging.

In terms of potential experimental implementation, fabrication is feasible using the same procedure as for a purely periodic structure~\cite{Zhang:2022-eabq4240:SCA}.
The dimensions and structure of the DBR and central gratings are within the fabrication resolution of e-beam lithography.
The precise focusing of the incident beam on the central region of the metasurface can also be achieved, for example an 8~$\mu$m diameter pump was used for second-harmonic generation from a similar structure~\cite{Yuan:2021-153901:PRL}, while SPDC was realized from just a single nano-resonator using an even tighter pump beam focusing~\cite{Marino:2019-1416:OPT, Zilli2023}. Importantly, our simulations show that for the designed metasurface cavity, the nonlinear rate is resilient with respect to variations in the pump beam size and position within the experimentally achievable precision.

We note further prospects to enhance the efficiency of SPDC. Recently, it was demonstrated that BICs with infinite Q-factors can exist in finite structures with perfectly reflective boundaries~\cite{liang2024boundstatescontinuuminfinite}, which can enable strong field localization.
There is also a potential to increase the efficiency by simultaneously engineering the resonances at the pump wavelength~\cite{Liu:2024-155424:PRB}.

We anticipate that this research will bring nonlinear metasurfaces closer to practical applications by revealing a new avenue for solving the critical problem of enhancing the photon-pair generation rate combined with photon state tunability.

\begin{acknowledgments}
This work was supported by the Australian Research Council (\url{https://doi.org/10.13039/501100000923}) Centre of Excellence for Transformative Meta-Optical Systems - TMOS (CE200100010)
\end{acknowledgments}

\section*{Data Availability Statement}
The data that support the findings of this study are available from the corresponding author upon reasonable request.

\appendix

\section{Simulations of eigen-wavelengths and reflection spectrum of DBR}

The metasurface cavity, consisting of the central grating and DBRs, was simulated using COMSOL Multiphysics. An eigenfrequency study was performed to identify the guided mode resonances supported by the periodic central grating structure. The period of the central gratings was initially tuned to achieve a resonance at 1570~nm. To simulate the angular dispersion of the eigen-wavelengths, the \( k_{x} \)-vector for Floquet periodicity in the central grating model was swept, as shown in Fig.~\ref{design_of_parameters_new}(c).

For the DBRs, illustrated in Fig.~\ref{design_of_parameters_new}(b), each unit cell comprises two regions. The effective refractive indices of these regions were directly extracted from COMSOL simulations. The unit cell dimensions were then analytically determined using the formula $p = {\lambda}/(4 n_{\mathrm{eff1}}) + {\lambda}/(4 n_{\mathrm{eff2}}) $ to ensure reflection within the desired spectral range. A combination of boundary mode analysis and a linear frequency-domain study was employed in COMSOL to simulate the reflection spectrum of the designed DBR, as shown in Fig.~\ref{design_of_parameters_new}(c). Similarly, to evaluate the angular dispersion of eigen-wavelengths in the full metasurface cavity model, the \( k_{z} \)-vector for Floquet periodicity was varied, with the results shown in Fig.~\ref{nonlinear_QNM}(a).





\bibliography{db_art_SPDC_DBR, manual_art_SPDC_DBR}

\begin{thebibliography}{52}%
\makeatletter
\providecommand \@ifxundefined [1]{%
 \@ifx{#1\undefined}
}%
\providecommand \@ifnum [1]{%
 \ifnum #1\expandafter \@firstoftwo
 \else \expandafter \@secondoftwo
 \fi
}%
\providecommand \@ifx [1]{%
 \ifx #1\expandafter \@firstoftwo
 \else \expandafter \@secondoftwo
 \fi
}%
\providecommand \natexlab [1]{#1}%
\providecommand \enquote  [1]{``#1''}%
\providecommand \bibnamefont  [1]{#1}%
\providecommand \bibfnamefont [1]{#1}%
\providecommand \citenamefont [1]{#1}%
\providecommand \href@noop [0]{\@secondoftwo}%
\providecommand \href [0]{\begingroup \@sanitize@url \@href}%
\providecommand \@href[1]{\@@startlink{#1}\@@href}%
\providecommand \@@href[1]{\endgroup#1\@@endlink}%
\providecommand \@sanitize@url [0]{\catcode `\\12\catcode `\$12\catcode `\&12\catcode `\#12\catcode `\^12\catcode `\_12\catcode `\%12\relax}%
\providecommand \@@startlink[1]{}%
\providecommand \@@endlink[0]{}%
\providecommand \url  [0]{\begingroup\@sanitize@url \@url }%
\providecommand \@url [1]{\endgroup\@href {#1}{\urlprefix }}%
\providecommand \urlprefix  [0]{URL }%
\providecommand \Eprint [0]{\href }%
\providecommand \doibase [0]{https://doi.org/}%
\providecommand \selectlanguage [0]{\@gobble}%
\providecommand \bibinfo  [0]{\@secondoftwo}%
\providecommand \bibfield  [0]{\@secondoftwo}%
\providecommand \translation [1]{[#1]}%
\providecommand \BibitemOpen [0]{}%
\providecommand \bibitemStop [0]{}%
\providecommand \bibitemNoStop [0]{.\EOS\space}%
\providecommand \EOS [0]{\spacefactor3000\relax}%
\providecommand \BibitemShut  [1]{\csname bibitem#1\endcsname}%
\let\auto@bib@innerbib\@empty
\bibitem [{\citenamefont {Slussarenko}\ and\ \citenamefont {Pryde}(2019)}]{Slussarenko:2019-41303:APPR}%
  \BibitemOpen
  \bibfield  {author} {\bibinfo {author} {\bibfnamefont {S.}~\bibnamefont {Slussarenko}}\ and\ \bibinfo {author} {\bibfnamefont {G.~J.}\ \bibnamefont {Pryde}},\ }\bibfield  {title} {{\selectlanguage {English}\bibinfo {title} {Photonic quantum information processing: A concise review}},\ }\href {https://doi.org/10.1063/1.5115814} {\bibfield  {journal} {\bibinfo  {journal} {Appl. Phys. Rev.}\ }\textbf {\bibinfo {volume} {6}},\ \bibinfo {pages} {041303} (\bibinfo {year} {2019})}\BibitemShut {NoStop}%
\bibitem [{\citenamefont {Clark}\ \emph {et~al.}(2021)\citenamefont {Clark}, \citenamefont {Chekhova}, \citenamefont {Matthews}, \citenamefont {Rarity},\ and\ \citenamefont {Oulton}}]{Clark:2021-60401:APL}%
  \BibitemOpen
  \bibfield  {author} {\bibinfo {author} {\bibfnamefont {A.~S.}\ \bibnamefont {Clark}}, \bibinfo {author} {\bibfnamefont {M.}~\bibnamefont {Chekhova}}, \bibinfo {author} {\bibfnamefont {J.~C.~F.}\ \bibnamefont {Matthews}}, \bibinfo {author} {\bibfnamefont {J.~G.}\ \bibnamefont {Rarity}},\ and\ \bibinfo {author} {\bibfnamefont {R.~F.}\ \bibnamefont {Oulton}},\ }\bibfield  {title} {{\selectlanguage {English}\bibinfo {title} {Special topic: Quantum sensing with correlated light sources}},\ }\href {https://doi.org/10.1063/5.0041043} {\bibfield  {journal} {\bibinfo  {journal} {Appl. Phys. Lett.}\ }\textbf {\bibinfo {volume} {118}},\ \bibinfo {pages} {060401} (\bibinfo {year} {2021})}\BibitemShut {NoStop}%
\bibitem [{\citenamefont {Li}\ \emph {et~al.}(2004)\citenamefont {Li}, \citenamefont {Chen}, \citenamefont {Voss}, \citenamefont {Sharping},\ and\ \citenamefont {Kumar}}]{Li2004}%
  \BibitemOpen
  \bibfield  {author} {\bibinfo {author} {\bibfnamefont {X.}~\bibnamefont {Li}}, \bibinfo {author} {\bibfnamefont {J.}~\bibnamefont {Chen}}, \bibinfo {author} {\bibfnamefont {P.}~\bibnamefont {Voss}}, \bibinfo {author} {\bibfnamefont {J.}~\bibnamefont {Sharping}},\ and\ \bibinfo {author} {\bibfnamefont {P.}~\bibnamefont {Kumar}},\ }\bibfield  {title} {\bibinfo {title} {{All-fiber photon-pair source for quantum communications: Improved generation of correlated photons}},\ }\href {https://doi.org/10.1364/opex.12.003737} {\bibfield  {journal} {\bibinfo  {journal} {Optics Express}\ }\textbf {\bibinfo {volume} {12}},\ \bibinfo {pages} {3737} (\bibinfo {year} {2004})}\BibitemShut {NoStop}%
\bibitem [{\citenamefont {Klyshko}(1988)}]{Klyshko:1988:PhotonsNonlinear}%
  \BibitemOpen
  \bibfield  {author} {\bibinfo {author} {\bibfnamefont {D.}~\bibnamefont {Klyshko}},\ }\href {https://doi.org/10.1201/9780203743508} {\emph {\bibinfo {title} {{Photons and Nonlinear Optics}}}}\ (\bibinfo  {publisher} {Gordon and Breach},\ \bibinfo {address} {New York},\ \bibinfo {year} {1988})\BibitemShut {NoStop}%
\bibitem [{\citenamefont {Couteau}(2018)}]{Couteau:2018-291:CTMP}%
  \BibitemOpen
  \bibfield  {author} {\bibinfo {author} {\bibfnamefont {C.}~\bibnamefont {Couteau}},\ }\bibfield  {title} {{\selectlanguage {English}\bibinfo {title} {Spontaneous parametric down-conversion}},\ }\href {https://doi.org/10.1080/00107514.2018.1488463} {\bibfield  {journal} {\bibinfo  {journal} {Contemp. Phys.}\ }\textbf {\bibinfo {volume} {59}},\ \bibinfo {pages} {291} (\bibinfo {year} {2018})}\BibitemShut {NoStop}%
\bibitem [{\citenamefont {Wang}\ \emph {et~al.}(2020)\citenamefont {Wang}, \citenamefont {Sciarrino}, \citenamefont {Laing},\ and\ \citenamefont {Thompson}}]{Wang:2020-273:NPHOT}%
  \BibitemOpen
  \bibfield  {author} {\bibinfo {author} {\bibfnamefont {J.~W.}\ \bibnamefont {Wang}}, \bibinfo {author} {\bibfnamefont {F.}~\bibnamefont {Sciarrino}}, \bibinfo {author} {\bibfnamefont {A.}~\bibnamefont {Laing}},\ and\ \bibinfo {author} {\bibfnamefont {M.~G.}\ \bibnamefont {Thompson}},\ }\bibfield  {title} {{\selectlanguage {English}\bibinfo {title} {Integrated photonic quantum technologies}},\ }\href {https://doi.org/10.1038/s41566-019-0532-1} {\bibfield  {journal} {\bibinfo  {journal} {Nat. Photon.}\ }\textbf {\bibinfo {volume} {14}},\ \bibinfo {pages} {273} (\bibinfo {year} {2020})}\BibitemShut {NoStop}%
\bibitem [{\citenamefont {Solntsev}\ \emph {et~al.}(2021)\citenamefont {Solntsev}, \citenamefont {Agarwal},\ and\ \citenamefont {Kivshar}}]{Solntsev:2021-327:NPHOT}%
  \BibitemOpen
  \bibfield  {author} {\bibinfo {author} {\bibfnamefont {A.~S.}\ \bibnamefont {Solntsev}}, \bibinfo {author} {\bibfnamefont {G.~S.}\ \bibnamefont {Agarwal}},\ and\ \bibinfo {author} {\bibfnamefont {Y.~Y.}\ \bibnamefont {Kivshar}},\ }\bibfield  {title} {{\selectlanguage {English}\bibinfo {title} {Metasurfaces for quantum photonics}},\ }\href {https://doi.org/10.1038/s41566-021-00793-z} {\bibfield  {journal} {\bibinfo  {journal} {Nat. Photon.}\ }\textbf {\bibinfo {volume} {15}},\ \bibinfo {pages} {327} (\bibinfo {year} {2021})}\BibitemShut {NoStop}%
\bibitem [{\citenamefont {Kan}\ and\ \citenamefont {Bozhevolnyi}(2023)}]{Kan:2023-2202759:ADOM}%
  \BibitemOpen
  \bibfield  {author} {\bibinfo {author} {\bibfnamefont {Y.~H.}\ \bibnamefont {Kan}}\ and\ \bibinfo {author} {\bibfnamefont {S.~I.}\ \bibnamefont {Bozhevolnyi}},\ }\bibfield  {title} {{\selectlanguage {English}\bibinfo {title} {Advances in metaphotonics empowered single photon emission}},\ }\href {https://doi.org/10.1002/adom.202202759} {\bibfield  {journal} {\bibinfo  {journal} {Adv. Opt. Mater.}\ }\textbf {\bibinfo {volume} {11}},\ \bibinfo {pages} {2202759} (\bibinfo {year} {2023})}\BibitemShut {NoStop}%
\bibitem [{\citenamefont {Ma}\ \emph {et~al.}(2024)\citenamefont {Ma}, \citenamefont {Zhang}, \citenamefont {Horder}, \citenamefont {Sukhorukov}, \citenamefont {Toth}, \citenamefont {Neshev},\ and\ \citenamefont {Aharonovich}}]{Ma:2024-2313589:ADM}%
  \BibitemOpen
  \bibfield  {author} {\bibinfo {author} {\bibfnamefont {J.~Y.}\ \bibnamefont {Ma}}, \bibinfo {author} {\bibfnamefont {J.~H.}\ \bibnamefont {Zhang}}, \bibinfo {author} {\bibfnamefont {J.}~\bibnamefont {Horder}}, \bibinfo {author} {\bibfnamefont {A.~A.}\ \bibnamefont {Sukhorukov}}, \bibinfo {author} {\bibfnamefont {M.}~\bibnamefont {Toth}}, \bibinfo {author} {\bibfnamefont {D.~N.}\ \bibnamefont {Neshev}},\ and\ \bibinfo {author} {\bibfnamefont {I.}~\bibnamefont {Aharonovich}},\ }\bibfield  {title} {{\selectlanguage {English}\bibinfo {title} {Engineering quantum light sources with flat optics}},\ }\href {https://doi.org/10.1002/adma.202313589} {\bibfield  {journal} {\bibinfo  {journal} {Adv. Mater.}\ }\textbf {\bibinfo {volume} {36}},\ \bibinfo {pages} {2313589} (\bibinfo {year} {2024})}\BibitemShut {NoStop}%
\bibitem [{\citenamefont {Poddubny}\ \emph {et~al.}(2016)\citenamefont {Poddubny}, \citenamefont {Iorsh},\ and\ \citenamefont {Sukhorukov}}]{Poddubny:2016-123901:PRL}%
  \BibitemOpen
  \bibfield  {author} {\bibinfo {author} {\bibfnamefont {A.~N.}\ \bibnamefont {Poddubny}}, \bibinfo {author} {\bibfnamefont {I.~V.}\ \bibnamefont {Iorsh}},\ and\ \bibinfo {author} {\bibfnamefont {A.~A.}\ \bibnamefont {Sukhorukov}},\ }\bibfield  {title} {{\selectlanguage {English}\bibinfo {title} {Generation of photon-plasmon quantum states in nonlinear hyperbolic metamaterials}},\ }\href {https://doi.org/10.1103/PhysRevLett.117.123901} {\bibfield  {journal} {\bibinfo  {journal} {Phys. Rev. Lett.}\ }\textbf {\bibinfo {volume} {117}},\ \bibinfo {pages} {123901} (\bibinfo {year} {2016})}\BibitemShut {NoStop}%
\bibitem [{\citenamefont {Davoyan}\ and\ \citenamefont {Atwater}(2018)}]{Davoyan:2018-608:OPT}%
  \BibitemOpen
  \bibfield  {author} {\bibinfo {author} {\bibfnamefont {A.}~\bibnamefont {Davoyan}}\ and\ \bibinfo {author} {\bibfnamefont {H.}~\bibnamefont {Atwater}},\ }\bibfield  {title} {{\selectlanguage {English}\bibinfo {title} {Quantum nonlinear light emission in metamaterials: broadband purcell enhancement of parametric downconversion}},\ }\href {https://doi.org/10.1364/OPTICA.5.000608} {\bibfield  {journal} {\bibinfo  {journal} {Optica}\ }\textbf {\bibinfo {volume} {5}},\ \bibinfo {pages} {608} (\bibinfo {year} {2018})}\BibitemShut {NoStop}%
\bibitem [{\citenamefont {Saleh}\ \emph {et~al.}(2018)\citenamefont {Saleh}, \citenamefont {Vezzoli}, \citenamefont {Caspani}, \citenamefont {Branny}, \citenamefont {Kumar}, \citenamefont {Gerardot},\ and\ \citenamefont {Faccio}}]{Saleh:2018-3862:SRP}%
  \BibitemOpen
  \bibfield  {author} {\bibinfo {author} {\bibfnamefont {H.~D.}\ \bibnamefont {Saleh}}, \bibinfo {author} {\bibfnamefont {S.}~\bibnamefont {Vezzoli}}, \bibinfo {author} {\bibfnamefont {L.}~\bibnamefont {Caspani}}, \bibinfo {author} {\bibfnamefont {A.}~\bibnamefont {Branny}}, \bibinfo {author} {\bibfnamefont {S.}~\bibnamefont {Kumar}}, \bibinfo {author} {\bibfnamefont {B.~D.}\ \bibnamefont {Gerardot}},\ and\ \bibinfo {author} {\bibfnamefont {D.}~\bibnamefont {Faccio}},\ }\bibfield  {title} {{\selectlanguage {English}\bibinfo {title} {Towards spontaneous parametric down conversion from monolayer {{MoS}$_2$}}},\ }\href {https://doi.org/10.1038/s41598-018-22270-4} {\bibfield  {journal} {\bibinfo  {journal} {Sci. Rep.}\ }\textbf {\bibinfo {volume} {8}},\ \bibinfo {pages} {3862} (\bibinfo {year} {2018})}\BibitemShut {NoStop}%
\bibitem [{\citenamefont {Guo}\ \emph {et~al.}(2023)\citenamefont {Guo}, \citenamefont {Qi}, \citenamefont {Zhang}, \citenamefont {Gao}, \citenamefont {Hu}, \citenamefont {Zhou}, \citenamefont {Zang}, \citenamefont {Zhao}, \citenamefont {Wang}, \citenamefont {Yan}, \citenamefont {Xu}, \citenamefont {Wu}, \citenamefont {Eda}, \citenamefont {Xiao}, \citenamefont {Yang}, \citenamefont {Gou}, \citenamefont {Feng}, \citenamefont {Guo}, \citenamefont {Zhou}, \citenamefont {Ren}, \citenamefont {Qiu}, \citenamefont {Pennycook},\ and\ \citenamefont {Wee}}]{Guo:2023-53:NAT}%
  \BibitemOpen
  \bibfield  {author} {\bibinfo {author} {\bibfnamefont {Q.~B.}\ \bibnamefont {Guo}}, \bibinfo {author} {\bibfnamefont {X.~Z.}\ \bibnamefont {Qi}}, \bibinfo {author} {\bibfnamefont {L.~S.}\ \bibnamefont {Zhang}}, \bibinfo {author} {\bibfnamefont {M.}~\bibnamefont {Gao}}, \bibinfo {author} {\bibfnamefont {S.~L.}\ \bibnamefont {Hu}}, \bibinfo {author} {\bibfnamefont {W.~J.}\ \bibnamefont {Zhou}}, \bibinfo {author} {\bibfnamefont {W.~J.}\ \bibnamefont {Zang}}, \bibinfo {author} {\bibfnamefont {X.~X.}\ \bibnamefont {Zhao}}, \bibinfo {author} {\bibfnamefont {J.~Y.}\ \bibnamefont {Wang}}, \bibinfo {author} {\bibfnamefont {B.~M.}\ \bibnamefont {Yan}}, \bibinfo {author} {\bibfnamefont {M.~Q.}\ \bibnamefont {Xu}}, \bibinfo {author} {\bibfnamefont {Y.~K.}\ \bibnamefont {Wu}}, \bibinfo {author} {\bibfnamefont {G.}~\bibnamefont {Eda}}, \bibinfo {author} {\bibfnamefont {Z.~W.}\ \bibnamefont {Xiao}}, \bibinfo {author} {\bibfnamefont {S.~Y.~A.}\ \bibnamefont {Yang}}, \bibinfo {author} {\bibfnamefont {H.~Y.}\ \bibnamefont
  {Gou}}, \bibinfo {author} {\bibfnamefont {Y.~P.}\ \bibnamefont {Feng}}, \bibinfo {author} {\bibfnamefont {G.~C.}\ \bibnamefont {Guo}}, \bibinfo {author} {\bibfnamefont {W.}~\bibnamefont {Zhou}}, \bibinfo {author} {\bibfnamefont {X.~F.}\ \bibnamefont {Ren}}, \bibinfo {author} {\bibfnamefont {C.~W.}\ \bibnamefont {Qiu}}, \bibinfo {author} {\bibfnamefont {S.~J.}\ \bibnamefont {Pennycook}},\ and\ \bibinfo {author} {\bibfnamefont {A.~T.~S.}\ \bibnamefont {Wee}},\ }\bibfield  {title} {{\selectlanguage {English}\bibinfo {title} {Ultrathin quantum light source with van der waals {{NbOCl}$_2$} crystal}},\ }\href {https://doi.org/10.1038/s41586-022-05393-7} {\bibfield  {journal} {\bibinfo  {journal} {Nature}\ }\textbf {\bibinfo {volume} {613}},\ \bibinfo {pages} {53} (\bibinfo {year} {2023})}\BibitemShut {NoStop}%
\bibitem [{\citenamefont {Weissflog}\ \emph {et~al.}(2024{\natexlab{a}})\citenamefont {Weissflog}, \citenamefont {Fedotova}, \citenamefont {Tang}, \citenamefont {Santos}, \citenamefont {Laudert}, \citenamefont {Shinde}, \citenamefont {Abtahi}, \citenamefont {Afsharnia}, \citenamefont {Perez}, \citenamefont {Ritter}, \citenamefont {Qin}, \citenamefont {Janousek}, \citenamefont {Shradha}, \citenamefont {Staude}, \citenamefont {Saravi}, \citenamefont {Pertsch}, \citenamefont {Setzpfandt}, \citenamefont {Lu},\ and\ \citenamefont {Eilenberger}}]{Weissflog:2024-7600:NCOM}%
  \BibitemOpen
  \bibfield  {author} {\bibinfo {author} {\bibfnamefont {M.~A.}\ \bibnamefont {Weissflog}}, \bibinfo {author} {\bibfnamefont {A.}~\bibnamefont {Fedotova}}, \bibinfo {author} {\bibfnamefont {Y.~L.}\ \bibnamefont {Tang}}, \bibinfo {author} {\bibfnamefont {E.~A.}\ \bibnamefont {Santos}}, \bibinfo {author} {\bibfnamefont {B.}~\bibnamefont {Laudert}}, \bibinfo {author} {\bibfnamefont {S.}~\bibnamefont {Shinde}}, \bibinfo {author} {\bibfnamefont {F.}~\bibnamefont {Abtahi}}, \bibinfo {author} {\bibfnamefont {M.}~\bibnamefont {Afsharnia}}, \bibinfo {author} {\bibfnamefont {I.~P.}\ \bibnamefont {Perez}}, \bibinfo {author} {\bibfnamefont {S.}~\bibnamefont {Ritter}}, \bibinfo {author} {\bibfnamefont {H.}~\bibnamefont {Qin}}, \bibinfo {author} {\bibfnamefont {J.}~\bibnamefont {Janousek}}, \bibinfo {author} {\bibfnamefont {S.}~\bibnamefont {Shradha}}, \bibinfo {author} {\bibfnamefont {I.}~\bibnamefont {Staude}}, \bibinfo {author} {\bibfnamefont {S.}~\bibnamefont {Saravi}}, \bibinfo {author} {\bibfnamefont
  {T.}~\bibnamefont {Pertsch}}, \bibinfo {author} {\bibfnamefont {F.}~\bibnamefont {Setzpfandt}}, \bibinfo {author} {\bibfnamefont {Y.~R.}\ \bibnamefont {Lu}},\ and\ \bibinfo {author} {\bibfnamefont {F.}~\bibnamefont {Eilenberger}},\ }\bibfield  {title} {{\selectlanguage {English}\bibinfo {title} {A tunable transition metal dichalcogenide entangled photon-pair source}},\ }\href {https://doi.org/10.1038/s41467-024-51843-3} {\bibfield  {journal} {\bibinfo  {journal} {Nat. Commun.}\ }\textbf {\bibinfo {volume} {15}},\ \bibinfo {pages} {7600} (\bibinfo {year} {2024}{\natexlab{a}})}\BibitemShut {NoStop}%
\bibitem [{\citenamefont {Sultanov}\ \emph {et~al.}(2024)\citenamefont {Sultanov}, \citenamefont {Kavcic}, \citenamefont {Kokkinakis}, \citenamefont {Sebastián}, \citenamefont {Chekhova},\ and\ \citenamefont {Humar}}]{Sultanov:2024-294:NAT}%
  \BibitemOpen
  \bibfield  {author} {\bibinfo {author} {\bibfnamefont {V.}~\bibnamefont {Sultanov}}, \bibinfo {author} {\bibfnamefont {A.}~\bibnamefont {Kavcic}}, \bibinfo {author} {\bibfnamefont {E.}~\bibnamefont {Kokkinakis}}, \bibinfo {author} {\bibfnamefont {N.}~\bibnamefont {Sebastián}}, \bibinfo {author} {\bibfnamefont {M.~V.}\ \bibnamefont {Chekhova}},\ and\ \bibinfo {author} {\bibfnamefont {M.}~\bibnamefont {Humar}},\ }\bibfield  {title} {{\selectlanguage {English}\bibinfo {title} {Tunable entangled photon-pair generation in a liquid crystal}},\ }\href {https://doi.org/10.1038/s41586-024-07543-5} {\bibfield  {journal} {\bibinfo  {journal} {Nature}\ }\textbf {\bibinfo {volume} {631}},\ \bibinfo {pages} {294} (\bibinfo {year} {2024})}\BibitemShut {NoStop}%
\bibitem [{\citenamefont {Okoth}\ \emph {et~al.}(2019)\citenamefont {Okoth}, \citenamefont {Cavanna}, \citenamefont {Santiago-Cruz},\ and\ \citenamefont {Chekhova}}]{Okoth:2019-263602:PRL}%
  \BibitemOpen
  \bibfield  {author} {\bibinfo {author} {\bibfnamefont {C.}~\bibnamefont {Okoth}}, \bibinfo {author} {\bibfnamefont {A.}~\bibnamefont {Cavanna}}, \bibinfo {author} {\bibfnamefont {T.}~\bibnamefont {Santiago-Cruz}},\ and\ \bibinfo {author} {\bibfnamefont {M.~V.}\ \bibnamefont {Chekhova}},\ }\bibfield  {title} {{\selectlanguage {English}\bibinfo {title} {Microscale generation of entangled photons without momentum conservation}},\ }\href {https://doi.org/10.1103/PhysRevLett.123.263602} {\bibfield  {journal} {\bibinfo  {journal} {Phys. Rev. Lett.}\ }\textbf {\bibinfo {volume} {123}},\ \bibinfo {pages} {263602} (\bibinfo {year} {2019})}\BibitemShut {NoStop}%
\bibitem [{\citenamefont {Okoth}\ \emph {et~al.}(2020)\citenamefont {Okoth}, \citenamefont {Kovlakov}, \citenamefont {Bonsel}, \citenamefont {Cavanna}, \citenamefont {Straupe}, \citenamefont {Kulik},\ and\ \citenamefont {Chekhova}}]{Okoth:2020-11801:PRA}%
  \BibitemOpen
  \bibfield  {author} {\bibinfo {author} {\bibfnamefont {C.}~\bibnamefont {Okoth}}, \bibinfo {author} {\bibfnamefont {E.}~\bibnamefont {Kovlakov}}, \bibinfo {author} {\bibfnamefont {F.}~\bibnamefont {Bonsel}}, \bibinfo {author} {\bibfnamefont {A.}~\bibnamefont {Cavanna}}, \bibinfo {author} {\bibfnamefont {S.}~\bibnamefont {Straupe}}, \bibinfo {author} {\bibfnamefont {S.~P.}\ \bibnamefont {Kulik}},\ and\ \bibinfo {author} {\bibfnamefont {M.~V.}\ \bibnamefont {Chekhova}},\ }\bibfield  {title} {{\selectlanguage {English}\bibinfo {title} {Idealized {E}instein-{P}odolsky-{R}osen states from non-phase-matched parametric down-conversion}},\ }\href {https://doi.org/10.1103/PhysRevA.101.011801} {\bibfield  {journal} {\bibinfo  {journal} {Phys. Rev. A}\ }\textbf {\bibinfo {volume} {101}},\ \bibinfo {pages} {011801} (\bibinfo {year} {2020})}\BibitemShut {NoStop}%
\bibitem [{\citenamefont {Santiago-Cruz}\ \emph {et~al.}(2021{\natexlab{a}})\citenamefont {Santiago-Cruz}, \citenamefont {Sultanov}, \citenamefont {Zhang}, \citenamefont {Krivitsky},\ and\ \citenamefont {Chekhova}}]{Santiago-Cruz:2021-653:OL}%
  \BibitemOpen
  \bibfield  {author} {\bibinfo {author} {\bibfnamefont {T.}~\bibnamefont {Santiago-Cruz}}, \bibinfo {author} {\bibfnamefont {V.}~\bibnamefont {Sultanov}}, \bibinfo {author} {\bibfnamefont {H.~Z.}\ \bibnamefont {Zhang}}, \bibinfo {author} {\bibfnamefont {L.~A.}\ \bibnamefont {Krivitsky}},\ and\ \bibinfo {author} {\bibfnamefont {M.~V.}\ \bibnamefont {Chekhova}},\ }\bibfield  {title} {{\selectlanguage {English}\bibinfo {title} {Entangled photons from subwavelength nonlinear films}},\ }\href {https://doi.org/10.1364/OL.411176} {\bibfield  {journal} {\bibinfo  {journal} {Opt. Lett.}\ }\textbf {\bibinfo {volume} {46}},\ \bibinfo {pages} {653} (\bibinfo {year} {2021}{\natexlab{a}})}\BibitemShut {NoStop}%
\bibitem [{\citenamefont {Sultanov}\ \emph {et~al.}(2022)\citenamefont {Sultanov}, \citenamefont {Santiago-Cruz},\ and\ \citenamefont {Chekhova}}]{Sultanov:2022-3872:OL}%
  \BibitemOpen
  \bibfield  {author} {\bibinfo {author} {\bibfnamefont {V.}~\bibnamefont {Sultanov}}, \bibinfo {author} {\bibfnamefont {T.}~\bibnamefont {Santiago-Cruz}},\ and\ \bibinfo {author} {\bibfnamefont {M.~V.}\ \bibnamefont {Chekhova}},\ }\bibfield  {title} {{\selectlanguage {English}\bibinfo {title} {Flat-optics generation of broadband photon pairs with tunable polarization entanglement}},\ }\href {https://doi.org/10.1364/OL.458133} {\bibfield  {journal} {\bibinfo  {journal} {Opt. Lett.}\ }\textbf {\bibinfo {volume} {47}},\ \bibinfo {pages} {3872} (\bibinfo {year} {2022})}\BibitemShut {NoStop}%
\bibitem [{\citenamefont {Marino}\ \emph {et~al.}(2019)\citenamefont {Marino}, \citenamefont {Solntsev}, \citenamefont {Xu}, \citenamefont {Gili}, \citenamefont {Carletti}, \citenamefont {Poddubny}, \citenamefont {Rahmani}, \citenamefont {Smirnova}, \citenamefont {Chen}, \citenamefont {Lemaitre}, \citenamefont {Zhang}, \citenamefont {Zayats}, \citenamefont {De~Angelis}, \citenamefont {Leo}, \citenamefont {Sukhorukov},\ and\ \citenamefont {Neshev}}]{Marino:2019-1416:OPT}%
  \BibitemOpen
  \bibfield  {author} {\bibinfo {author} {\bibfnamefont {G.}~\bibnamefont {Marino}}, \bibinfo {author} {\bibfnamefont {A.~S.}\ \bibnamefont {Solntsev}}, \bibinfo {author} {\bibfnamefont {L.}~\bibnamefont {Xu}}, \bibinfo {author} {\bibfnamefont {V.~F.}\ \bibnamefont {Gili}}, \bibinfo {author} {\bibfnamefont {L.}~\bibnamefont {Carletti}}, \bibinfo {author} {\bibfnamefont {A.~N.}\ \bibnamefont {Poddubny}}, \bibinfo {author} {\bibfnamefont {M.}~\bibnamefont {Rahmani}}, \bibinfo {author} {\bibfnamefont {D.~A.}\ \bibnamefont {Smirnova}}, \bibinfo {author} {\bibfnamefont {H.~T.}\ \bibnamefont {Chen}}, \bibinfo {author} {\bibfnamefont {A.}~\bibnamefont {Lemaitre}}, \bibinfo {author} {\bibfnamefont {G.~Q.}\ \bibnamefont {Zhang}}, \bibinfo {author} {\bibfnamefont {A.~V.}\ \bibnamefont {Zayats}}, \bibinfo {author} {\bibfnamefont {C.}~\bibnamefont {De~Angelis}}, \bibinfo {author} {\bibfnamefont {G.}~\bibnamefont {Leo}}, \bibinfo {author} {\bibfnamefont {A.~A.}\ \bibnamefont {Sukhorukov}},\ and\ \bibinfo {author}
  {\bibfnamefont {D.~N.}\ \bibnamefont {Neshev}},\ }\bibfield  {title} {{\selectlanguage {English}\bibinfo {title} {Spontaneous photon-pair generation from a dielectric nanoantenna}},\ }\href {https://doi.org/10.1364/OPTICA.6.001416} {\bibfield  {journal} {\bibinfo  {journal} {Optica}\ }\textbf {\bibinfo {volume} {6}},\ \bibinfo {pages} {1416} (\bibinfo {year} {2019})}\BibitemShut {NoStop}%
\bibitem [{\citenamefont {Nikolaeva}\ \emph {et~al.}(2021)\citenamefont {Nikolaeva}, \citenamefont {Frizyuk}, \citenamefont {Olekhno}, \citenamefont {Solntsev},\ and\ \citenamefont {Petrov}}]{Nikolaeva:2021-43703:PRA}%
  \BibitemOpen
  \bibfield  {author} {\bibinfo {author} {\bibfnamefont {A.}~\bibnamefont {Nikolaeva}}, \bibinfo {author} {\bibfnamefont {K.}~\bibnamefont {Frizyuk}}, \bibinfo {author} {\bibfnamefont {N.}~\bibnamefont {Olekhno}}, \bibinfo {author} {\bibfnamefont {A.}~\bibnamefont {Solntsev}},\ and\ \bibinfo {author} {\bibfnamefont {M.}~\bibnamefont {Petrov}},\ }\bibfield  {title} {{\selectlanguage {English}\bibinfo {title} {Directional emission of down-converted photons from a dielectric nanoresonator}},\ }\href {https://doi.org/10.1103/PhysRevA.103.043703} {\bibfield  {journal} {\bibinfo  {journal} {Phys. Rev. A}\ }\textbf {\bibinfo {volume} {103}},\ \bibinfo {pages} {043703} (\bibinfo {year} {2021})}\BibitemShut {NoStop}%
\bibitem [{\citenamefont {Duong}\ \emph {et~al.}(2022)\citenamefont {Duong}, \citenamefont {Saerens}, \citenamefont {Timpu}, \citenamefont {Buscaglia}, \citenamefont {Buscaglia}, \citenamefont {Morandi}, \citenamefont {Muller}, \citenamefont {Maeder}, \citenamefont {Kaufmann}, \citenamefont {Solntsev},\ and\ \citenamefont {Grange}}]{Duong:2022-3696:OME}%
  \BibitemOpen
  \bibfield  {author} {\bibinfo {author} {\bibfnamefont {N.~M.~H.}\ \bibnamefont {Duong}}, \bibinfo {author} {\bibfnamefont {G.}~\bibnamefont {Saerens}}, \bibinfo {author} {\bibfnamefont {F.}~\bibnamefont {Timpu}}, \bibinfo {author} {\bibfnamefont {M.~T.}\ \bibnamefont {Buscaglia}}, \bibinfo {author} {\bibfnamefont {V.}~\bibnamefont {Buscaglia}}, \bibinfo {author} {\bibfnamefont {A.}~\bibnamefont {Morandi}}, \bibinfo {author} {\bibfnamefont {J.~S.}\ \bibnamefont {Muller}}, \bibinfo {author} {\bibfnamefont {A.}~\bibnamefont {Maeder}}, \bibinfo {author} {\bibfnamefont {F.}~\bibnamefont {Kaufmann}}, \bibinfo {author} {\bibfnamefont {A.~S.}\ \bibnamefont {Solntsev}},\ and\ \bibinfo {author} {\bibfnamefont {R.}~\bibnamefont {Grange}},\ }\bibfield  {title} {{\selectlanguage {English}\bibinfo {title} {Spontaneous parametric down-conversion in bottom-up grown lithium niobate microcubes}},\ }\href {https://doi.org/10.1364/OME.462981} {\bibfield  {journal} {\bibinfo  {journal} {Opt. Mater. Express}\ }\textbf {\bibinfo
  {volume} {12}},\ \bibinfo {pages} {3696} (\bibinfo {year} {2022})}\BibitemShut {NoStop}%
\bibitem [{\citenamefont {Zilli}\ \emph {et~al.}(2023)\citenamefont {Zilli}, \citenamefont {Sultanov}, \citenamefont {Poloczek}, \citenamefont {Ferrera}, \citenamefont {Luan}, \citenamefont {Kokkinakis}, \citenamefont {Santiago-Cruz}, \citenamefont {Carletti}, \citenamefont {Finazzi}, \citenamefont {Toma}, \citenamefont {Chekhova},\ and\ \citenamefont {Celebrano}}]{Zilli2023}%
  \BibitemOpen
  \bibfield  {author} {\bibinfo {author} {\bibfnamefont {A.}~\bibnamefont {Zilli}}, \bibinfo {author} {\bibfnamefont {V.}~\bibnamefont {Sultanov}}, \bibinfo {author} {\bibfnamefont {M.}~\bibnamefont {Poloczek}}, \bibinfo {author} {\bibfnamefont {M.}~\bibnamefont {Ferrera}}, \bibinfo {author} {\bibfnamefont {Y.}~\bibnamefont {Luan}}, \bibinfo {author} {\bibfnamefont {E.~T.}\ \bibnamefont {Kokkinakis}}, \bibinfo {author} {\bibfnamefont {T.}~\bibnamefont {Santiago-Cruz}}, \bibinfo {author} {\bibfnamefont {L.}~\bibnamefont {Carletti}}, \bibinfo {author} {\bibfnamefont {M.}~\bibnamefont {Finazzi}}, \bibinfo {author} {\bibfnamefont {A.}~\bibnamefont {Toma}}, \bibinfo {author} {\bibfnamefont {M.~V.}\ \bibnamefont {Chekhova}},\ and\ \bibinfo {author} {\bibfnamefont {M.}~\bibnamefont {Celebrano}},\ }\bibfield  {title} {\bibinfo {title} {{Spontaneous Parametric Down-Conversion Beaming from a Lithium Niobate Nanostructured Resonator}},\ }\href {https://doi.org/10.1364/CLEO_FS.2023.ftu4c.8} {\bibfield  {journal}
  {\bibinfo  {journal} {CLEO: Fundamental Science}\ ,\ \bibinfo {pages} {FTu4C.8}} (\bibinfo {year} {2023})}\BibitemShut {NoStop}%
\bibitem [{\citenamefont {Saerens}\ \emph {et~al.}(2023)\citenamefont {Saerens}, \citenamefont {Dursap}, \citenamefont {Hesner}, \citenamefont {Duong}, \citenamefont {Solntsev}, \citenamefont {Morandi}, \citenamefont {Maeder}, \citenamefont {Karvounis}, \citenamefont {Regreny}, \citenamefont {Chapman}, \citenamefont {Danescu}, \citenamefont {Chauvin}, \citenamefont {Penuelas},\ and\ \citenamefont {Grange}}]{Saerens:2023-3245:NANL}%
  \BibitemOpen
  \bibfield  {author} {\bibinfo {author} {\bibfnamefont {G.}~\bibnamefont {Saerens}}, \bibinfo {author} {\bibfnamefont {T.}~\bibnamefont {Dursap}}, \bibinfo {author} {\bibfnamefont {I.}~\bibnamefont {Hesner}}, \bibinfo {author} {\bibfnamefont {N.~M.~H.}\ \bibnamefont {Duong}}, \bibinfo {author} {\bibfnamefont {A.~S.}\ \bibnamefont {Solntsev}}, \bibinfo {author} {\bibfnamefont {A.}~\bibnamefont {Morandi}}, \bibinfo {author} {\bibfnamefont {A.}~\bibnamefont {Maeder}}, \bibinfo {author} {\bibfnamefont {A.}~\bibnamefont {Karvounis}}, \bibinfo {author} {\bibfnamefont {P.}~\bibnamefont {Regreny}}, \bibinfo {author} {\bibfnamefont {R.~J.}\ \bibnamefont {Chapman}}, \bibinfo {author} {\bibfnamefont {A.}~\bibnamefont {Danescu}}, \bibinfo {author} {\bibfnamefont {N.}~\bibnamefont {Chauvin}}, \bibinfo {author} {\bibfnamefont {J.}~\bibnamefont {Penuelas}},\ and\ \bibinfo {author} {\bibfnamefont {R.}~\bibnamefont {Grange}},\ }\bibfield  {title} {{\selectlanguage {English}\bibinfo {title} {Background-free near-infrared
  biphoton emission from single {GaAs} nanowires}},\ }\href {https://doi.org/10.1021/acs.nanolett.3c00026} {\bibfield  {journal} {\bibinfo  {journal} {Nano Lett.}\ }\textbf {\bibinfo {volume} {23}},\ \bibinfo {pages} {3245} (\bibinfo {year} {2023})}\BibitemShut {NoStop}%
\bibitem [{\citenamefont {Olekhno}\ \emph {et~al.}(2024)\citenamefont {Olekhno}, \citenamefont {Petrov}, \citenamefont {Iorsh}, \citenamefont {Sukhorukov},\ and\ \citenamefont {Solntsev}}]{Olekhno:2024-245416:PRB}%
  \BibitemOpen
  \bibfield  {author} {\bibinfo {author} {\bibfnamefont {N.~A.}\ \bibnamefont {Olekhno}}, \bibinfo {author} {\bibfnamefont {M.~I.}\ \bibnamefont {Petrov}}, \bibinfo {author} {\bibfnamefont {I.}~\bibnamefont {Iorsh}}, \bibinfo {author} {\bibfnamefont {A.~A.}\ \bibnamefont {Sukhorukov}},\ and\ \bibinfo {author} {\bibfnamefont {A.~S.}\ \bibnamefont {Solntsev}},\ }\bibfield  {title} {{\selectlanguage {English}\bibinfo {title} {{Generating N00N states of surface plasmon polaritons with $N=2$ by a single nanoparticle}}},\ }\href {https://doi.org/10.1103/PhysRevB.109.245416} {\bibfield  {journal} {\bibinfo  {journal} {Phys. Rev. B}\ }\textbf {\bibinfo {volume} {109}},\ \bibinfo {pages} {245416} (\bibinfo {year} {2024})}\BibitemShut {NoStop}%
\bibitem [{\citenamefont {Weissflog}\ \emph {et~al.}(2024{\natexlab{b}})\citenamefont {Weissflog}, \citenamefont {Dezert}, \citenamefont {Vinel}, \citenamefont {Gigli}, \citenamefont {Leo}, \citenamefont {Pertsch}, \citenamefont {Setzpfandt}, \citenamefont {Borne},\ and\ \citenamefont {Saravi}}]{Weissflog:2024-11403:APPR}%
  \BibitemOpen
  \bibfield  {author} {\bibinfo {author} {\bibfnamefont {M.~A.}\ \bibnamefont {Weissflog}}, \bibinfo {author} {\bibfnamefont {R.}~\bibnamefont {Dezert}}, \bibinfo {author} {\bibfnamefont {V.}~\bibnamefont {Vinel}}, \bibinfo {author} {\bibfnamefont {C.}~\bibnamefont {Gigli}}, \bibinfo {author} {\bibfnamefont {G.}~\bibnamefont {Leo}}, \bibinfo {author} {\bibfnamefont {T.}~\bibnamefont {Pertsch}}, \bibinfo {author} {\bibfnamefont {F.}~\bibnamefont {Setzpfandt}}, \bibinfo {author} {\bibfnamefont {A.}~\bibnamefont {Borne}},\ and\ \bibinfo {author} {\bibfnamefont {S.}~\bibnamefont {Saravi}},\ }\bibfield  {title} {{\selectlanguage {English}\bibinfo {title} {Nonlinear nanoresonators for {B}ell state generation}},\ }\href {https://doi.org/10.1063/5.0172240} {\bibfield  {journal} {\bibinfo  {journal} {Appl. Phys. Rev.}\ }\textbf {\bibinfo {volume} {11}},\ \bibinfo {pages} {011403} (\bibinfo {year} {2024}{\natexlab{b}})}\BibitemShut {NoStop}%
\bibitem [{\citenamefont {Kuznetsov}\ \emph {et~al.}(2024)\citenamefont {Kuznetsov}, \citenamefont {Brongersma}, \citenamefont {Yao}, \citenamefont {Chen}, \citenamefont {Levy}, \citenamefont {Tsai}, \citenamefont {Zheludev}, \citenamefont {Faraon}, \citenamefont {Arbabi}, \citenamefont {Yu}, \citenamefont {Chanda}, \citenamefont {Crozier}, \citenamefont {Kildishev}, \citenamefont {Wang}, \citenamefont {Yang}, \citenamefont {Valentine}, \citenamefont {Genevet}, \citenamefont {Fan}, \citenamefont {Miller}, \citenamefont {Majumdar}, \citenamefont {Fröch}, \citenamefont {Brady}, \citenamefont {Heide}, \citenamefont {Veeraraghavan}, \citenamefont {Engheta}, \citenamefont {Alù}, \citenamefont {Polman}, \citenamefont {Atwater}, \citenamefont {Thureja}, \citenamefont {Paniagua-Dominguez}, \citenamefont {Ha}, \citenamefont {Barreda}, \citenamefont {Schuller}, \citenamefont {Staude}, \citenamefont {Grinblat}, \citenamefont {Kivshar}, \citenamefont {Peana}, \citenamefont {Yelin}, \citenamefont {Senichev},
  \citenamefont {Shalaev}, \citenamefont {Saha}, \citenamefont {Boltasseva}, \citenamefont {Rho}, \citenamefont {Oh}, \citenamefont {Kim}, \citenamefont {Park}, \citenamefont {Devlin},\ and\ \citenamefont {Pala}}]{Kuznetsov:2024-816:ACSP}%
  \BibitemOpen
  \bibfield  {author} {\bibinfo {author} {\bibfnamefont {A.~I.}\ \bibnamefont {Kuznetsov}}, \bibinfo {author} {\bibfnamefont {M.~L.}\ \bibnamefont {Brongersma}}, \bibinfo {author} {\bibfnamefont {J.}~\bibnamefont {Yao}}, \bibinfo {author} {\bibfnamefont {M.~K.}\ \bibnamefont {Chen}}, \bibinfo {author} {\bibfnamefont {U.}~\bibnamefont {Levy}}, \bibinfo {author} {\bibfnamefont {D.~P.}\ \bibnamefont {Tsai}}, \bibinfo {author} {\bibfnamefont {N.~I.}\ \bibnamefont {Zheludev}}, \bibinfo {author} {\bibfnamefont {A.}~\bibnamefont {Faraon}}, \bibinfo {author} {\bibfnamefont {A.}~\bibnamefont {Arbabi}}, \bibinfo {author} {\bibfnamefont {N.~F.}\ \bibnamefont {Yu}}, \bibinfo {author} {\bibfnamefont {D.}~\bibnamefont {Chanda}}, \bibinfo {author} {\bibfnamefont {K.~B.}\ \bibnamefont {Crozier}}, \bibinfo {author} {\bibfnamefont {A.~V.}\ \bibnamefont {Kildishev}}, \bibinfo {author} {\bibfnamefont {H.}~\bibnamefont {Wang}}, \bibinfo {author} {\bibfnamefont {J.~K.~W.}\ \bibnamefont {Yang}}, \bibinfo {author} {\bibfnamefont
  {J.~G.}\ \bibnamefont {Valentine}}, \bibinfo {author} {\bibfnamefont {P.}~\bibnamefont {Genevet}}, \bibinfo {author} {\bibfnamefont {J.~A.}\ \bibnamefont {Fan}}, \bibinfo {author} {\bibfnamefont {O.~D.}\ \bibnamefont {Miller}}, \bibinfo {author} {\bibfnamefont {A.}~\bibnamefont {Majumdar}}, \bibinfo {author} {\bibfnamefont {J.~E.}\ \bibnamefont {Fröch}}, \bibinfo {author} {\bibfnamefont {D.}~\bibnamefont {Brady}}, \bibinfo {author} {\bibfnamefont {F.}~\bibnamefont {Heide}}, \bibinfo {author} {\bibfnamefont {A.}~\bibnamefont {Veeraraghavan}}, \bibinfo {author} {\bibfnamefont {N.}~\bibnamefont {Engheta}}, \bibinfo {author} {\bibfnamefont {A.}~\bibnamefont {Alù}}, \bibinfo {author} {\bibfnamefont {A.}~\bibnamefont {Polman}}, \bibinfo {author} {\bibfnamefont {H.~A.}\ \bibnamefont {Atwater}}, \bibinfo {author} {\bibfnamefont {P.}~\bibnamefont {Thureja}}, \bibinfo {author} {\bibfnamefont {R.}~\bibnamefont {Paniagua-Dominguez}}, \bibinfo {author} {\bibfnamefont {S.~T.}\ \bibnamefont {Ha}}, \bibinfo {author}
  {\bibfnamefont {A.~I.}\ \bibnamefont {Barreda}}, \bibinfo {author} {\bibfnamefont {J.~A.}\ \bibnamefont {Schuller}}, \bibinfo {author} {\bibfnamefont {I.}~\bibnamefont {Staude}}, \bibinfo {author} {\bibfnamefont {G.}~\bibnamefont {Grinblat}}, \bibinfo {author} {\bibfnamefont {Y.}~\bibnamefont {Kivshar}}, \bibinfo {author} {\bibfnamefont {S.}~\bibnamefont {Peana}}, \bibinfo {author} {\bibfnamefont {S.~F.}\ \bibnamefont {Yelin}}, \bibinfo {author} {\bibfnamefont {A.}~\bibnamefont {Senichev}}, \bibinfo {author} {\bibfnamefont {V.~M.}\ \bibnamefont {Shalaev}}, \bibinfo {author} {\bibfnamefont {S.}~\bibnamefont {Saha}}, \bibinfo {author} {\bibfnamefont {A.}~\bibnamefont {Boltasseva}}, \bibinfo {author} {\bibfnamefont {J.}~\bibnamefont {Rho}}, \bibinfo {author} {\bibfnamefont {D.~K.}\ \bibnamefont {Oh}}, \bibinfo {author} {\bibfnamefont {J.}~\bibnamefont {Kim}}, \bibinfo {author} {\bibfnamefont {J.}~\bibnamefont {Park}}, \bibinfo {author} {\bibfnamefont {R.}~\bibnamefont {Devlin}},\ and\ \bibinfo {author}
  {\bibfnamefont {R.~A.}\ \bibnamefont {Pala}},\ }\bibfield  {title} {{\selectlanguage {English}\bibinfo {title} {Roadmap for optical metasurfaces}},\ }\href {https://doi.org/10.1021/acsphotonics.3c00457} {\bibfield  {journal} {\bibinfo  {journal} {ACS Photonics}\ }\textbf {\bibinfo {volume} {11}},\ \bibinfo {pages} {816} (\bibinfo {year} {2024})}\BibitemShut {NoStop}%
\bibitem [{\citenamefont {Schulz}\ \emph {et~al.}(2024)\citenamefont {Schulz}, \citenamefont {Oulton}, \citenamefont {Kenney}, \citenamefont {Alù}, \citenamefont {Staude}, \citenamefont {Bashiri}, \citenamefont {Fedorova}, \citenamefont {Kolkowski}, \citenamefont {Koenderink}, \citenamefont {Xiao}, \citenamefont {Yang}, \citenamefont {Peveler}, \citenamefont {Clark}, \citenamefont {Perrakis}, \citenamefont {Tasolamprou}, \citenamefont {Kafesaki}, \citenamefont {Zaleska}, \citenamefont {Dickson}, \citenamefont {Richards}, \citenamefont {Zayats}, \citenamefont {Ren}, \citenamefont {Kivshar}, \citenamefont {Maier}, \citenamefont {Chen}, \citenamefont {Ansari}, \citenamefont {Gan}, \citenamefont {Alexeev}, \citenamefont {Krauss}, \citenamefont {Di~Falco}, \citenamefont {Gennaro}, \citenamefont {Santiago-Cruz}, \citenamefont {Brener}, \citenamefont {Chekhova}, \citenamefont {Ma}, \citenamefont {Vogler-Neuling}, \citenamefont {Weigand}, \citenamefont {Talts}, \citenamefont {Occhiodori}, \citenamefont {Grange},
  \citenamefont {Rahmani}, \citenamefont {Xu}, \citenamefont {Kamali}, \citenamefont {Arababi}, \citenamefont {Faraon}, \citenamefont {Harwood}, \citenamefont {Vezzoli}, \citenamefont {Sapienza}, \citenamefont {Lalanne}, \citenamefont {Dmitriev}, \citenamefont {Rockstuhl}, \citenamefont {Sprafke}, \citenamefont {Vynck}, \citenamefont {Upham}, \citenamefont {Alam}, \citenamefont {De~Leon}, \citenamefont {Boyd}, \citenamefont {Padilla}, \citenamefont {Malof}, \citenamefont {Jana}, \citenamefont {Yang}, \citenamefont {Colom}, \citenamefont {Song}, \citenamefont {Genevet}, \citenamefont {Achouri}, \citenamefont {Evlyukhin}, \citenamefont {Lemmer},\ and\ \citenamefont {Fernandez-Corbaton}}]{Schulz:2024-260701:APL}%
  \BibitemOpen
  \bibfield  {author} {\bibinfo {author} {\bibfnamefont {S.~A.}\ \bibnamefont {Schulz}}, \bibinfo {author} {\bibfnamefont {R.~F.}\ \bibnamefont {Oulton}}, \bibinfo {author} {\bibfnamefont {M.}~\bibnamefont {Kenney}}, \bibinfo {author} {\bibfnamefont {A.}~\bibnamefont {Alù}}, \bibinfo {author} {\bibfnamefont {I.}~\bibnamefont {Staude}}, \bibinfo {author} {\bibfnamefont {A.}~\bibnamefont {Bashiri}}, \bibinfo {author} {\bibfnamefont {Z.}~\bibnamefont {Fedorova}}, \bibinfo {author} {\bibfnamefont {R.}~\bibnamefont {Kolkowski}}, \bibinfo {author} {\bibfnamefont {A.~F.}\ \bibnamefont {Koenderink}}, \bibinfo {author} {\bibfnamefont {X.}~\bibnamefont {Xiao}}, \bibinfo {author} {\bibfnamefont {J.}~\bibnamefont {Yang}}, \bibinfo {author} {\bibfnamefont {W.~J.}\ \bibnamefont {Peveler}}, \bibinfo {author} {\bibfnamefont {A.~W.}\ \bibnamefont {Clark}}, \bibinfo {author} {\bibfnamefont {G.}~\bibnamefont {Perrakis}}, \bibinfo {author} {\bibfnamefont {A.~C.}\ \bibnamefont {Tasolamprou}}, \bibinfo {author} {\bibfnamefont
  {M.}~\bibnamefont {Kafesaki}}, \bibinfo {author} {\bibfnamefont {A.}~\bibnamefont {Zaleska}}, \bibinfo {author} {\bibfnamefont {W.}~\bibnamefont {Dickson}}, \bibinfo {author} {\bibfnamefont {D.}~\bibnamefont {Richards}}, \bibinfo {author} {\bibfnamefont {A.}~\bibnamefont {Zayats}}, \bibinfo {author} {\bibfnamefont {H.}~\bibnamefont {Ren}}, \bibinfo {author} {\bibfnamefont {Y.}~\bibnamefont {Kivshar}}, \bibinfo {author} {\bibfnamefont {S.}~\bibnamefont {Maier}}, \bibinfo {author} {\bibfnamefont {X.}~\bibnamefont {Chen}}, \bibinfo {author} {\bibfnamefont {M.~A.}\ \bibnamefont {Ansari}}, \bibinfo {author} {\bibfnamefont {Y.}~\bibnamefont {Gan}}, \bibinfo {author} {\bibfnamefont {A.}~\bibnamefont {Alexeev}}, \bibinfo {author} {\bibfnamefont {T.~F.}\ \bibnamefont {Krauss}}, \bibinfo {author} {\bibfnamefont {A.}~\bibnamefont {Di~Falco}}, \bibinfo {author} {\bibfnamefont {S.~D.}\ \bibnamefont {Gennaro}}, \bibinfo {author} {\bibfnamefont {T.}~\bibnamefont {Santiago-Cruz}}, \bibinfo {author} {\bibfnamefont
  {I.}~\bibnamefont {Brener}}, \bibinfo {author} {\bibfnamefont {M.~V.}\ \bibnamefont {Chekhova}}, \bibinfo {author} {\bibfnamefont {R.-M.}\ \bibnamefont {Ma}}, \bibinfo {author} {\bibfnamefont {V.~V.}\ \bibnamefont {Vogler-Neuling}}, \bibinfo {author} {\bibfnamefont {H.~C.}\ \bibnamefont {Weigand}}, \bibinfo {author} {\bibfnamefont {U.-L.}\ \bibnamefont {Talts}}, \bibinfo {author} {\bibfnamefont {I.}~\bibnamefont {Occhiodori}}, \bibinfo {author} {\bibfnamefont {R.}~\bibnamefont {Grange}}, \bibinfo {author} {\bibfnamefont {M.}~\bibnamefont {Rahmani}}, \bibinfo {author} {\bibfnamefont {L.}~\bibnamefont {Xu}}, \bibinfo {author} {\bibfnamefont {S.~M.}\ \bibnamefont {Kamali}}, \bibinfo {author} {\bibfnamefont {E.}~\bibnamefont {Arababi}}, \bibinfo {author} {\bibfnamefont {A.}~\bibnamefont {Faraon}}, \bibinfo {author} {\bibfnamefont {A.~C.}\ \bibnamefont {Harwood}}, \bibinfo {author} {\bibfnamefont {S.}~\bibnamefont {Vezzoli}}, \bibinfo {author} {\bibfnamefont {R.}~\bibnamefont {Sapienza}}, \bibinfo {author}
  {\bibfnamefont {P.}~\bibnamefont {Lalanne}}, \bibinfo {author} {\bibfnamefont {A.}~\bibnamefont {Dmitriev}}, \bibinfo {author} {\bibfnamefont {C.}~\bibnamefont {Rockstuhl}}, \bibinfo {author} {\bibfnamefont {A.}~\bibnamefont {Sprafke}}, \bibinfo {author} {\bibfnamefont {K.}~\bibnamefont {Vynck}}, \bibinfo {author} {\bibfnamefont {J.}~\bibnamefont {Upham}}, \bibinfo {author} {\bibfnamefont {M.~Z.}\ \bibnamefont {Alam}}, \bibinfo {author} {\bibfnamefont {I.}~\bibnamefont {De~Leon}}, \bibinfo {author} {\bibfnamefont {R.~W.}\ \bibnamefont {Boyd}}, \bibinfo {author} {\bibfnamefont {W.~J.}\ \bibnamefont {Padilla}}, \bibinfo {author} {\bibfnamefont {J.~M.}\ \bibnamefont {Malof}}, \bibinfo {author} {\bibfnamefont {A.}~\bibnamefont {Jana}}, \bibinfo {author} {\bibfnamefont {Z.}~\bibnamefont {Yang}}, \bibinfo {author} {\bibfnamefont {R.}~\bibnamefont {Colom}}, \bibinfo {author} {\bibfnamefont {Q.}~\bibnamefont {Song}}, \bibinfo {author} {\bibfnamefont {P.}~\bibnamefont {Genevet}}, \bibinfo {author} {\bibfnamefont
  {K.}~\bibnamefont {Achouri}}, \bibinfo {author} {\bibfnamefont {A.~B.}\ \bibnamefont {Evlyukhin}}, \bibinfo {author} {\bibfnamefont {U.}~\bibnamefont {Lemmer}},\ and\ \bibinfo {author} {\bibfnamefont {I.}~\bibnamefont {Fernandez-Corbaton}},\ }\bibfield  {title} {\bibinfo {title} {Roadmap on photonic metasurfaces},\ }\href {https://doi.org/10.1063/5.0204694} {\bibfield  {journal} {\bibinfo  {journal} {Appl. Phys. Lett.}\ }\textbf {\bibinfo {volume} {124}},\ \bibinfo {pages} {260701} (\bibinfo {year} {2024})}\BibitemShut {NoStop}%
\bibitem [{\citenamefont {Santiago-Cruz}\ \emph {et~al.}(2021{\natexlab{b}})\citenamefont {Santiago-Cruz}, \citenamefont {Fedotova}, \citenamefont {Sultanov}, \citenamefont {Weissflog}, \citenamefont {Arslan}, \citenamefont {Younesi}, \citenamefont {Pertsch}, \citenamefont {Staude}, \citenamefont {Setzpfandt},\ and\ \citenamefont {Chekhova}}]{Santiago-Cruz:2021-4423:NANL}%
  \BibitemOpen
  \bibfield  {author} {\bibinfo {author} {\bibfnamefont {T.}~\bibnamefont {Santiago-Cruz}}, \bibinfo {author} {\bibfnamefont {A.}~\bibnamefont {Fedotova}}, \bibinfo {author} {\bibfnamefont {V.}~\bibnamefont {Sultanov}}, \bibinfo {author} {\bibfnamefont {M.~A.}\ \bibnamefont {Weissflog}}, \bibinfo {author} {\bibfnamefont {D.}~\bibnamefont {Arslan}}, \bibinfo {author} {\bibfnamefont {M.}~\bibnamefont {Younesi}}, \bibinfo {author} {\bibfnamefont {T.}~\bibnamefont {Pertsch}}, \bibinfo {author} {\bibfnamefont {I.}~\bibnamefont {Staude}}, \bibinfo {author} {\bibfnamefont {F.}~\bibnamefont {Setzpfandt}},\ and\ \bibinfo {author} {\bibfnamefont {M.}~\bibnamefont {Chekhova}},\ }\bibfield  {title} {{\selectlanguage {English}\bibinfo {title} {Photon pairs from resonant metasurfaces}},\ }\href {https://doi.org/10.1021/acs.nanolett.1c01125} {\bibfield  {journal} {\bibinfo  {journal} {Nano Lett.}\ }\textbf {\bibinfo {volume} {21}},\ \bibinfo {pages} {4423} (\bibinfo {year} {2021}{\natexlab{b}})}\BibitemShut {NoStop}%
\bibitem [{\citenamefont {Zhang}\ \emph {et~al.}(2022)\citenamefont {Zhang}, \citenamefont {Ma}, \citenamefont {Parry}, \citenamefont {Cai}, \citenamefont {Camacho-Morales}, \citenamefont {Xu}, \citenamefont {Neshev},\ and\ \citenamefont {Sukhorukov}}]{Zhang:2022-eabq4240:SCA}%
  \BibitemOpen
  \bibfield  {author} {\bibinfo {author} {\bibfnamefont {J.~H.}\ \bibnamefont {Zhang}}, \bibinfo {author} {\bibfnamefont {J.~Y.}\ \bibnamefont {Ma}}, \bibinfo {author} {\bibfnamefont {M.}~\bibnamefont {Parry}}, \bibinfo {author} {\bibfnamefont {M.}~\bibnamefont {Cai}}, \bibinfo {author} {\bibfnamefont {R.}~\bibnamefont {Camacho-Morales}}, \bibinfo {author} {\bibfnamefont {L.}~\bibnamefont {Xu}}, \bibinfo {author} {\bibfnamefont {D.~N.}\ \bibnamefont {Neshev}},\ and\ \bibinfo {author} {\bibfnamefont {A.~A.}\ \bibnamefont {Sukhorukov}},\ }\bibfield  {title} {{\selectlanguage {English}\bibinfo {title} {Spatially entangled photon pairs from lithium niobate nonlocal metasurfaces}},\ }\href {https://doi.org/10.1126/sciadv.abq4240} {\bibfield  {journal} {\bibinfo  {journal} {Sci. Adv.}\ }\textbf {\bibinfo {volume} {8}},\ \bibinfo {pages} {eabq4240} (\bibinfo {year} {2022})}\BibitemShut {NoStop}%
\bibitem [{\citenamefont {Santiago-Cruz}\ \emph {et~al.}(2022)\citenamefont {Santiago-Cruz}, \citenamefont {Gennaro}, \citenamefont {Mitrofanov}, \citenamefont {Addamane}, \citenamefont {Reno}, \citenamefont {Brener},\ and\ \citenamefont {Chekhova}}]{Santiago-Cruz:2022-991:SCI}%
  \BibitemOpen
  \bibfield  {author} {\bibinfo {author} {\bibfnamefont {T.}~\bibnamefont {Santiago-Cruz}}, \bibinfo {author} {\bibfnamefont {S.~D.}\ \bibnamefont {Gennaro}}, \bibinfo {author} {\bibfnamefont {O.}~\bibnamefont {Mitrofanov}}, \bibinfo {author} {\bibfnamefont {S.}~\bibnamefont {Addamane}}, \bibinfo {author} {\bibfnamefont {J.}~\bibnamefont {Reno}}, \bibinfo {author} {\bibfnamefont {I.}~\bibnamefont {Brener}},\ and\ \bibinfo {author} {\bibfnamefont {M.~V.}\ \bibnamefont {Chekhova}},\ }\bibfield  {title} {{\selectlanguage {English}\bibinfo {title} {Resonant metasurfaces for generating complex quantum states}},\ }\href {https://doi.org/10.1126/science.abq8684} {\bibfield  {journal} {\bibinfo  {journal} {Science}\ }\textbf {\bibinfo {volume} {377}},\ \bibinfo {pages} {991} (\bibinfo {year} {2022})}\BibitemShut {NoStop}%
\bibitem [{\citenamefont {Ma}\ \emph {et~al.}(2023)\citenamefont {Ma}, \citenamefont {Zhang}, \citenamefont {Jiang}, \citenamefont {Fan}, \citenamefont {Parry}, \citenamefont {Neshev},\ and\ \citenamefont {Sukhorukov}}]{Ma:2023-8091:NANL}%
  \BibitemOpen
  \bibfield  {author} {\bibinfo {author} {\bibfnamefont {J.}~\bibnamefont {Ma}}, \bibinfo {author} {\bibfnamefont {J.}~\bibnamefont {Zhang}}, \bibinfo {author} {\bibfnamefont {Y.}~\bibnamefont {Jiang}}, \bibinfo {author} {\bibfnamefont {T.}~\bibnamefont {Fan}}, \bibinfo {author} {\bibfnamefont {M.}~\bibnamefont {Parry}}, \bibinfo {author} {\bibfnamefont {D.~N.}\ \bibnamefont {Neshev}},\ and\ \bibinfo {author} {\bibfnamefont {A.~A.}\ \bibnamefont {Sukhorukov}},\ }\bibfield  {title} {{\selectlanguage {English}\bibinfo {title} {Polarization engineering of entangled photons from a lithium niobate nonlinear metasurface}},\ }\href {https://doi.org/10.1021/acs.nanolett.3c02055} {\bibfield  {journal} {\bibinfo  {journal} {Nano Lett.}\ }\textbf {\bibinfo {volume} {23}},\ \bibinfo {pages} {8091} (\bibinfo {year} {2023})}\BibitemShut {NoStop}%
\bibitem [{\citenamefont {Jia}\ \emph {et~al.}(2025)\citenamefont {Jia}, \citenamefont {Saerens}, \citenamefont {Talts}, \citenamefont {Weigand}, \citenamefont {Chapman}, \citenamefont {Li}, \citenamefont {Grange},\ and\ \citenamefont {Yang}}]{Jia:2025-eads3576:SCA}%
  \BibitemOpen
  \bibfield  {author} {\bibinfo {author} {\bibfnamefont {W.~H.}\ \bibnamefont {Jia}}, \bibinfo {author} {\bibfnamefont {G.}~\bibnamefont {Saerens}}, \bibinfo {author} {\bibfnamefont {U.}~\bibnamefont {Talts}}, \bibinfo {author} {\bibfnamefont {H.}~\bibnamefont {Weigand}}, \bibinfo {author} {\bibfnamefont {R.~J.}\ \bibnamefont {Chapman}}, \bibinfo {author} {\bibfnamefont {L.}~\bibnamefont {Li}}, \bibinfo {author} {\bibfnamefont {R.}~\bibnamefont {Grange}},\ and\ \bibinfo {author} {\bibfnamefont {Y.~M.}\ \bibnamefont {Yang}},\ }\bibfield  {title} {{\selectlanguage {English}\bibinfo {title} {Polarization-entangled {B}ell state generation from an epsilon-near-zero metasurface}},\ }\href {https://doi.org/10.1126/sciadv.ads3576} {\bibfield  {journal} {\bibinfo  {journal} {Sci. Adv.}\ }\textbf {\bibinfo {volume} {11}},\ \bibinfo {pages} {eads3576} (\bibinfo {year} {2025})}\BibitemShut {NoStop}%
\bibitem [{\citenamefont {Weissflog}\ \emph {et~al.}(2024{\natexlab{c}})\citenamefont {Weissflog}, \citenamefont {Ma}, \citenamefont {Zhang}, \citenamefont {Fan}, \citenamefont {Lung}, \citenamefont {Pertsch}, \citenamefont {Neshev}, \citenamefont {Saravi}, \citenamefont {Setzpfandt},\ and\ \citenamefont {Sukhorukov}}]{Weissflog:2024-3563:NANP}%
  \BibitemOpen
  \bibfield  {author} {\bibinfo {author} {\bibfnamefont {M.~A.}\ \bibnamefont {Weissflog}}, \bibinfo {author} {\bibfnamefont {J.~Y.}\ \bibnamefont {Ma}}, \bibinfo {author} {\bibfnamefont {J.~H.}\ \bibnamefont {Zhang}}, \bibinfo {author} {\bibfnamefont {T.~M.}\ \bibnamefont {Fan}}, \bibinfo {author} {\bibfnamefont {S.}~\bibnamefont {Lung}}, \bibinfo {author} {\bibfnamefont {T.}~\bibnamefont {Pertsch}}, \bibinfo {author} {\bibfnamefont {D.~N.}\ \bibnamefont {Neshev}}, \bibinfo {author} {\bibfnamefont {S.}~\bibnamefont {Saravi}}, \bibinfo {author} {\bibfnamefont {F.}~\bibnamefont {Setzpfandt}},\ and\ \bibinfo {author} {\bibfnamefont {A.~A.}\ \bibnamefont {Sukhorukov}},\ }\bibfield  {title} {{\selectlanguage {English}\bibinfo {title} {Directionally tunable co- and counterpropagating photon pairs from a nonlinear metasurface}},\ }\href {https://doi.org/10.1515/nanoph-2024-0122} {\bibfield  {journal} {\bibinfo  {journal} {Nanophotonics}\ }\textbf {\bibinfo {volume} {13}},\ \bibinfo {pages} {3563} (\bibinfo {year}
  {2024}{\natexlab{c}})}\BibitemShut {NoStop}%
\bibitem [{\citenamefont {Yuan}\ \emph {et~al.}(2021)\citenamefont {Yuan}, \citenamefont {Wu}, \citenamefont {Dang}, \citenamefont {Zeng}, \citenamefont {Qi}, \citenamefont {Guo}, \citenamefont {Ren},\ and\ \citenamefont {Xia}}]{Yuan:2021-153901:PRL}%
  \BibitemOpen
  \bibfield  {author} {\bibinfo {author} {\bibfnamefont {S.}~\bibnamefont {Yuan}}, \bibinfo {author} {\bibfnamefont {Y.~K.}\ \bibnamefont {Wu}}, \bibinfo {author} {\bibfnamefont {Z.~Z.}\ \bibnamefont {Dang}}, \bibinfo {author} {\bibfnamefont {C.}~\bibnamefont {Zeng}}, \bibinfo {author} {\bibfnamefont {X.~Z.}\ \bibnamefont {Qi}}, \bibinfo {author} {\bibfnamefont {G.~C.}\ \bibnamefont {Guo}}, \bibinfo {author} {\bibfnamefont {X.~F.}\ \bibnamefont {Ren}},\ and\ \bibinfo {author} {\bibfnamefont {J.~S.}\ \bibnamefont {Xia}},\ }\bibfield  {title} {{\selectlanguage {English}\bibinfo {title} {Strongly enhanced second harmonic generation in a thin film lithium niobate heterostructure cavity}},\ }\href {https://doi.org/10.1103/PhysRevLett.127.153901} {\bibfield  {journal} {\bibinfo  {journal} {Phys. Rev. Lett.}\ }\textbf {\bibinfo {volume} {127}},\ \bibinfo {pages} {153901} (\bibinfo {year} {2021})}\BibitemShut {NoStop}%
\bibitem [{\citenamefont {Parry}\ \emph {et~al.}(2021)\citenamefont {Parry}, \citenamefont {Mazzanti}, \citenamefont {Poddubny}, \citenamefont {Della~Valle}, \citenamefont {Neshev},\ and\ \citenamefont {Sukhorukov}}]{Parry:2021-55001:ADP}%
  \BibitemOpen
  \bibfield  {author} {\bibinfo {author} {\bibfnamefont {M.}~\bibnamefont {Parry}}, \bibinfo {author} {\bibfnamefont {A.}~\bibnamefont {Mazzanti}}, \bibinfo {author} {\bibfnamefont {A.}~\bibnamefont {Poddubny}}, \bibinfo {author} {\bibfnamefont {G.}~\bibnamefont {Della~Valle}}, \bibinfo {author} {\bibfnamefont {D.~N.}\ \bibnamefont {Neshev}},\ and\ \bibinfo {author} {\bibfnamefont {A.~A.}\ \bibnamefont {Sukhorukov}},\ }\bibfield  {title} {{\selectlanguage {English}\bibinfo {title} {Enhanced generation of nondegenerate photon pairs in nonlinear metasurfaces}},\ }\href {https://doi.org/10.1117/1.AP.3.5.055001} {\bibfield  {journal} {\bibinfo  {journal} {Adv. Photon.}\ }\textbf {\bibinfo {volume} {3}},\ \bibinfo {pages} {055001} (\bibinfo {year} {2021})}\BibitemShut {NoStop}%
\bibitem [{\citenamefont {Mazzanti}\ \emph {et~al.}(2022)\citenamefont {Mazzanti}, \citenamefont {Parry}, \citenamefont {Poddubny}, \citenamefont {Della~Valle}, \citenamefont {Neshev},\ and\ \citenamefont {Sukhorukov}}]{Mazzanti:2022-35006:NJP}%
  \BibitemOpen
  \bibfield  {author} {\bibinfo {author} {\bibfnamefont {A.}~\bibnamefont {Mazzanti}}, \bibinfo {author} {\bibfnamefont {M.}~\bibnamefont {Parry}}, \bibinfo {author} {\bibfnamefont {A.~N.}\ \bibnamefont {Poddubny}}, \bibinfo {author} {\bibfnamefont {G.}~\bibnamefont {Della~Valle}}, \bibinfo {author} {\bibfnamefont {D.~N.}\ \bibnamefont {Neshev}},\ and\ \bibinfo {author} {\bibfnamefont {A.~A.}\ \bibnamefont {Sukhorukov}},\ }\bibfield  {title} {{\selectlanguage {English}\bibinfo {title} {Enhanced generation of angle correlated photon-pairs in nonlinear metasurfaces}},\ }\href {https://doi.org/10.1088/1367-2630/ac599e} {\bibfield  {journal} {\bibinfo  {journal} {New J. Phys.}\ }\textbf {\bibinfo {volume} {24}},\ \bibinfo {pages} {035006} (\bibinfo {year} {2022})}\BibitemShut {NoStop}%
\bibitem [{\citenamefont {Liu}\ \emph {et~al.}(2024)\citenamefont {Liu}, \citenamefont {Qin}, \citenamefont {Feng}, \citenamefont {Tu}, \citenamefont {Guo}, \citenamefont {Wu},\ and\ \citenamefont {Xiao}}]{Liu:2024-155424:PRB}%
  \BibitemOpen
  \bibfield  {author} {\bibinfo {author} {\bibfnamefont {T.}~\bibnamefont {Liu}}, \bibinfo {author} {\bibfnamefont {M.}~\bibnamefont {Qin}}, \bibinfo {author} {\bibfnamefont {S.}~\bibnamefont {Feng}}, \bibinfo {author} {\bibfnamefont {X.}~\bibnamefont {Tu}}, \bibinfo {author} {\bibfnamefont {T.}~\bibnamefont {Guo}}, \bibinfo {author} {\bibfnamefont {F.}~\bibnamefont {Wu}},\ and\ \bibinfo {author} {\bibfnamefont {S.}~\bibnamefont {Xiao}},\ }\bibfield  {title} {\bibinfo {title} {Efficient photon-pair generation empowered by dual quasibound states in the continuum},\ }\href {https://doi.org/10.1103/physrevb.109.155424} {\bibfield  {journal} {\bibinfo  {journal} {Phys. Rev. B}\ }\textbf {\bibinfo {volume} {109}},\ \bibinfo {pages} {155424} (\bibinfo {year} {2024})}\BibitemShut {NoStop}%
\bibitem [{\citenamefont {Saravi}\ \emph {et~al.}(2021)\citenamefont {Saravi}, \citenamefont {Pertsch},\ and\ \citenamefont {Setzpfandt}}]{Saravi:2021-2100789:ADOM}%
  \BibitemOpen
  \bibfield  {author} {\bibinfo {author} {\bibfnamefont {S.}~\bibnamefont {Saravi}}, \bibinfo {author} {\bibfnamefont {T.}~\bibnamefont {Pertsch}},\ and\ \bibinfo {author} {\bibfnamefont {F.}~\bibnamefont {Setzpfandt}},\ }\bibfield  {title} {{\selectlanguage {English}\bibinfo {title} {Lithium niobate on insulator: An emerging platform for integrated quantum photonics}},\ }\href {https://doi.org/10.1002/adom.202100789} {\bibfield  {journal} {\bibinfo  {journal} {Adv. Opt. Mater.}\ }\textbf {\bibinfo {volume} {9}},\ \bibinfo {pages} {2100789} (\bibinfo {year} {2021})}\BibitemShut {NoStop}%
\bibitem [{\citenamefont {Lee}\ \emph {et~al.}(2020)\citenamefont {Lee}, \citenamefont {Lee}, \citenamefont {Oh}, \citenamefont {Badloe}, \citenamefont {Ok},\ and\ \citenamefont {Rho}}]{Lee:2020-4108:SENS}%
  \BibitemOpen
  \bibfield  {author} {\bibinfo {author} {\bibfnamefont {T.}~\bibnamefont {Lee}}, \bibinfo {author} {\bibfnamefont {C.}~\bibnamefont {Lee}}, \bibinfo {author} {\bibfnamefont {D.~K.}\ \bibnamefont {Oh}}, \bibinfo {author} {\bibfnamefont {T.}~\bibnamefont {Badloe}}, \bibinfo {author} {\bibfnamefont {J.~G.}\ \bibnamefont {Ok}},\ and\ \bibinfo {author} {\bibfnamefont {J.}~\bibnamefont {Rho}},\ }\bibfield  {title} {{\selectlanguage {English}\bibinfo {title} {Scalable and high-throughput top-down manufacturing of optical metasurfaces}},\ }\href {https://doi.org/10.3390/s20154108} {\bibfield  {journal} {\bibinfo  {journal} {Sensors}\ }\textbf {\bibinfo {volume} {20}},\ \bibinfo {pages} {4108} (\bibinfo {year} {2020})}\BibitemShut {NoStop}%
\bibitem [{\citenamefont {Sheppard}(1995)}]{Sheppard1995}%
  \BibitemOpen
  \bibfield  {author} {\bibinfo {author} {\bibfnamefont {C.~J.}\ \bibnamefont {Sheppard}},\ }\bibfield  {title} {\bibinfo {title} {{Approximate calculation of the reflection coefficient from a stratified medium}},\ }\href {https://doi.org/10.1088/0963-9659/4/5/018} {\bibfield  {journal} {\bibinfo  {journal} {Pure Appl. Opt.}\ }\textbf {\bibinfo {volume} {4}},\ \bibinfo {pages} {665} (\bibinfo {year} {1995})}\BibitemShut {NoStop}%
\bibitem [{\citenamefont {Lai}\ \emph {et~al.}(1990)\citenamefont {Lai}, \citenamefont {Leung}, \citenamefont {Young}, \citenamefont {Barber},\ and\ \citenamefont {Hill}}]{Lai:1990-5187:PRA}%
  \BibitemOpen
  \bibfield  {author} {\bibinfo {author} {\bibfnamefont {H.~M.}\ \bibnamefont {Lai}}, \bibinfo {author} {\bibfnamefont {P.~T.}\ \bibnamefont {Leung}}, \bibinfo {author} {\bibfnamefont {K.}~\bibnamefont {Young}}, \bibinfo {author} {\bibfnamefont {P.~W.}\ \bibnamefont {Barber}},\ and\ \bibinfo {author} {\bibfnamefont {S.~C.}\ \bibnamefont {Hill}},\ }\bibfield  {title} {{\selectlanguage {English}\bibinfo {title} {Time-independent perturbation for leaking electromagnetic modes in open systems with application to resonances in microdroplets}},\ }\href {https://doi.org/10.1103/PhysRevA.41.5187} {\bibfield  {journal} {\bibinfo  {journal} {Phys. Rev. A}\ }\textbf {\bibinfo {volume} {41}},\ \bibinfo {pages} {5187} (\bibinfo {year} {1990})}\BibitemShut {NoStop}%
\bibitem [{\citenamefont {Lalanne}\ \emph {et~al.}(2018)\citenamefont {Lalanne}, \citenamefont {Yan}, \citenamefont {Vynck}, \citenamefont {Sauvan},\ and\ \citenamefont {Hugonin}}]{Lalanne:2018-1700113:LPR}%
  \BibitemOpen
  \bibfield  {author} {\bibinfo {author} {\bibfnamefont {P.}~\bibnamefont {Lalanne}}, \bibinfo {author} {\bibfnamefont {W.}~\bibnamefont {Yan}}, \bibinfo {author} {\bibfnamefont {K.}~\bibnamefont {Vynck}}, \bibinfo {author} {\bibfnamefont {C.}~\bibnamefont {Sauvan}},\ and\ \bibinfo {author} {\bibfnamefont {J.~P.}\ \bibnamefont {Hugonin}},\ }\bibfield  {title} {{\selectlanguage {English}\bibinfo {title} {Light interaction with photonic and plasmonic resonances}},\ }\href {https://doi.org/10.1002/lpor.201700113} {\bibfield  {journal} {\bibinfo  {journal} {Laser Photon. Rev.}\ }\textbf {\bibinfo {volume} {12}},\ \bibinfo {pages} {1700113} (\bibinfo {year} {2018})}\BibitemShut {NoStop}%
\bibitem [{\citenamefont {Sauvan}\ \emph {et~al.}(2013)\citenamefont {Sauvan}, \citenamefont {Hugonin}, \citenamefont {Maksymov},\ and\ \citenamefont {Lalanne}}]{Sauvan:2013-237401:PRL}%
  \BibitemOpen
  \bibfield  {author} {\bibinfo {author} {\bibfnamefont {C.}~\bibnamefont {Sauvan}}, \bibinfo {author} {\bibfnamefont {J.~P.}\ \bibnamefont {Hugonin}}, \bibinfo {author} {\bibfnamefont {I.~S.}\ \bibnamefont {Maksymov}},\ and\ \bibinfo {author} {\bibfnamefont {P.}~\bibnamefont {Lalanne}},\ }\bibfield  {title} {{\selectlanguage {English}\bibinfo {title} {Theory of the spontaneous optical emission of nanosize photonic and plasmon resonators}},\ }\href {https://doi.org/10.1103/PhysRevLett.110.237401} {\bibfield  {journal} {\bibinfo  {journal} {Phys. Rev. Lett.}\ }\textbf {\bibinfo {volume} {110}},\ \bibinfo {pages} {237401} (\bibinfo {year} {2013})}\BibitemShut {NoStop}%
\bibitem [{\citenamefont {Gigli}\ \emph {et~al.}(2020)\citenamefont {Gigli}, \citenamefont {Wu}, \citenamefont {Marino}, \citenamefont {Borne}, \citenamefont {Leo},\ and\ \citenamefont {Lalanne}}]{Gigli:2020-1197:ACSP}%
  \BibitemOpen
  \bibfield  {author} {\bibinfo {author} {\bibfnamefont {C.}~\bibnamefont {Gigli}}, \bibinfo {author} {\bibfnamefont {T.}~\bibnamefont {Wu}}, \bibinfo {author} {\bibfnamefont {G.}~\bibnamefont {Marino}}, \bibinfo {author} {\bibfnamefont {A.}~\bibnamefont {Borne}}, \bibinfo {author} {\bibfnamefont {G.}~\bibnamefont {Leo}},\ and\ \bibinfo {author} {\bibfnamefont {P.}~\bibnamefont {Lalanne}},\ }\bibfield  {title} {{\selectlanguage {English}\bibinfo {title} {Quasinormal-mode non-{H}ermitian modeling and design in nonlinear nano-optics}},\ }\href {https://doi.org/10.1021/acsphotonics.0c00014} {\bibfield  {journal} {\bibinfo  {journal} {ACS Photonics}\ }\textbf {\bibinfo {volume} {7}},\ \bibinfo {pages} {1197} (\bibinfo {year} {2020})}\BibitemShut {NoStop}%
\bibitem [{\citenamefont {Shukhin}\ \emph {et~al.}(2024)\citenamefont {Shukhin}, \citenamefont {Hurvitz}, \citenamefont {Trajtenberg-mills}, \citenamefont {Arie},\ and\ \citenamefont {Eisenberg}}]{Shukhin:2024-10158:OE}%
  \BibitemOpen
  \bibfield  {author} {\bibinfo {author} {\bibfnamefont {A.}~\bibnamefont {Shukhin}}, \bibinfo {author} {\bibfnamefont {I.}~\bibnamefont {Hurvitz}}, \bibinfo {author} {\bibfnamefont {S.}~\bibnamefont {Trajtenberg-mills}}, \bibinfo {author} {\bibfnamefont {A.}~\bibnamefont {Arie}},\ and\ \bibinfo {author} {\bibfnamefont {H.}~\bibnamefont {Eisenberg}},\ }\bibfield  {title} {{\selectlanguage {English}\bibinfo {title} {Two-dimensional control of a biphoton joint spectrum}},\ }\href {https://doi.org/10.1364/OE.510497} {\bibfield  {journal} {\bibinfo  {journal} {Opt. Express}\ }\textbf {\bibinfo {volume} {32}},\ \bibinfo {pages} {10158} (\bibinfo {year} {2024})}\BibitemShut {NoStop}%
\bibitem [{\citenamefont {Pittman}\ \emph {et~al.}(1995)\citenamefont {Pittman}, \citenamefont {Shih}, \citenamefont {Strekalov},\ and\ \citenamefont {Sergienko}}]{Pittman:1995-3429:PRA}%
  \BibitemOpen
  \bibfield  {author} {\bibinfo {author} {\bibfnamefont {T.~B.}\ \bibnamefont {Pittman}}, \bibinfo {author} {\bibfnamefont {Y.~H.}\ \bibnamefont {Shih}}, \bibinfo {author} {\bibfnamefont {D.~V.}\ \bibnamefont {Strekalov}},\ and\ \bibinfo {author} {\bibfnamefont {A.~V.}\ \bibnamefont {Sergienko}},\ }\bibfield  {title} {{\selectlanguage {English}\bibinfo {title} {Optical imaging by means of 2-photon quantum entanglement}},\ }\href {https://doi.org/10.1103/PhysRevA.52.R3429} {\bibfield  {journal} {\bibinfo  {journal} {Phys. Rev. A}\ }\textbf {\bibinfo {volume} {52}},\ \bibinfo {pages} {R3429} (\bibinfo {year} {1995})}\BibitemShut {NoStop}%
\bibitem [{\citenamefont {Moreau}\ \emph {et~al.}(2019)\citenamefont {Moreau}, \citenamefont {Toninelli}, \citenamefont {Gregory},\ and\ \citenamefont {Padgett}}]{Moreau:2019-367:NRP}%
  \BibitemOpen
  \bibfield  {author} {\bibinfo {author} {\bibfnamefont {P.~A.}\ \bibnamefont {Moreau}}, \bibinfo {author} {\bibfnamefont {E.}~\bibnamefont {Toninelli}}, \bibinfo {author} {\bibfnamefont {T.}~\bibnamefont {Gregory}},\ and\ \bibinfo {author} {\bibfnamefont {M.~J.}\ \bibnamefont {Padgett}},\ }\bibfield  {title} {{\selectlanguage {English}\bibinfo {title} {Imaging with quantum states of light}},\ }\href {https://doi.org/10.1038/s42254-019-0056-0} {\bibfield  {journal} {\bibinfo  {journal} {Nat. Rev. Phys.}\ }\textbf {\bibinfo {volume} {1}},\ \bibinfo {pages} {367} (\bibinfo {year} {2019})}\BibitemShut {NoStop}%
\bibitem [{\citenamefont {Ma}\ \emph {et~al.}(2025)\citenamefont {Ma}, \citenamefont {Ren}, \citenamefont {Zhang}, \citenamefont {Meng}, \citenamefont {McManus-Barrett}, \citenamefont {Crozier},\ and\ \citenamefont {Sukhorukov}}]{Ma:2025-2:ELI}%
  \BibitemOpen
  \bibfield  {author} {\bibinfo {author} {\bibfnamefont {J.~Y.}\ \bibnamefont {Ma}}, \bibinfo {author} {\bibfnamefont {J.~L.}\ \bibnamefont {Ren}}, \bibinfo {author} {\bibfnamefont {J.~H.}\ \bibnamefont {Zhang}}, \bibinfo {author} {\bibfnamefont {J.~J.}\ \bibnamefont {Meng}}, \bibinfo {author} {\bibfnamefont {C.}~\bibnamefont {McManus-Barrett}}, \bibinfo {author} {\bibfnamefont {K.~B.}\ \bibnamefont {Crozier}},\ and\ \bibinfo {author} {\bibfnamefont {A.~A.}\ \bibnamefont {Sukhorukov}},\ }\bibfield  {title} {{\selectlanguage {English}\bibinfo {title} {Quantum imaging using spatially entangled photon pairs from a nonlinear metasurface}},\ }\href {https://doi.org/10.1186/s43593-024-00080-8} {\bibfield  {journal} {\bibinfo  {journal} {eLight}\ }\textbf {\bibinfo {volume} {5}},\ \bibinfo {pages} {2} (\bibinfo {year} {2025})}\BibitemShut {NoStop}%
\bibitem [{\citenamefont {Griffiths}\ \emph {et~al.}(2023)\citenamefont {Griffiths}, \citenamefont {San~Lo}, \citenamefont {Dynes}, \citenamefont {Woodward},\ and\ \citenamefont {Shields}}]{Griffiths:2023-44040:PRAP}%
  \BibitemOpen
  \bibfield  {author} {\bibinfo {author} {\bibfnamefont {B.}~\bibnamefont {Griffiths}}, \bibinfo {author} {\bibfnamefont {Y.}~\bibnamefont {San~Lo}}, \bibinfo {author} {\bibfnamefont {J.~F.}\ \bibnamefont {Dynes}}, \bibinfo {author} {\bibfnamefont {R.~I.}\ \bibnamefont {Woodward}},\ and\ \bibinfo {author} {\bibfnamefont {A.~J.}\ \bibnamefont {Shields}},\ }\bibfield  {title} {{\selectlanguage {English}\bibinfo {title} {Optical transmitter tunable over a 65-nm wavelength range around 1550 nm for quantum key distribution}},\ }\href {https://doi.org/10.1103/PhysRevApplied.20.044040} {\bibfield  {journal} {\bibinfo  {journal} {Phys. Rev. Appl.}\ }\textbf {\bibinfo {volume} {20}},\ \bibinfo {pages} {044040} (\bibinfo {year} {2023})}\BibitemShut {NoStop}%
\bibitem [{\citenamefont {Jin}\ \emph {et~al.}(2021)\citenamefont {Jin}, \citenamefont {Mishra},\ and\ \citenamefont {Argyropoulos}}]{Jin:2021-19903:NASC}%
  \BibitemOpen
  \bibfield  {author} {\bibinfo {author} {\bibfnamefont {B.~Y.}\ \bibnamefont {Jin}}, \bibinfo {author} {\bibfnamefont {D.}~\bibnamefont {Mishra}},\ and\ \bibinfo {author} {\bibfnamefont {C.}~\bibnamefont {Argyropoulos}},\ }\bibfield  {title} {{\selectlanguage {English}\bibinfo {title} {Efficient single-photon pair generation by spontaneous parametric down-conversion in nonlinear plasmonic metasurfaces}},\ }\href {https://doi.org/10.1039/d1nr05379e} {\bibfield  {journal} {\bibinfo  {journal} {Nanoscale}\ }\textbf {\bibinfo {volume} {13}},\ \bibinfo {pages} {19903} (\bibinfo {year} {2021})}\BibitemShut {NoStop}%
\bibitem [{\citenamefont {Liang}\ \emph {et~al.}(2024)\citenamefont {Liang}, \citenamefont {Liu}, \citenamefont {Zeng}, \citenamefont {Cai},\ and\ \citenamefont {Ning}}]{liang2024boundstatescontinuuminfinite}%
  \BibitemOpen
  \bibfield  {author} {\bibinfo {author} {\bibfnamefont {H.}~\bibnamefont {Liang}}, \bibinfo {author} {\bibfnamefont {Y.}~\bibnamefont {Liu}}, \bibinfo {author} {\bibfnamefont {Y.-J.}\ \bibnamefont {Zeng}}, \bibinfo {author} {\bibfnamefont {Y.}~\bibnamefont {Cai}},\ and\ \bibinfo {author} {\bibfnamefont {T.}~\bibnamefont {Ning}},\ }\href {https://doi.org/10.48550/arXiv.2411.12374} {\bibinfo {title} {Bound states in the continuum of infinite quality factor in finite unit cells}} (\bibinfo {year} {2024}),\ \Eprint {https://arxiv.org/abs/2411.12374} {arXiv:2411.12374 [physics.optics]} \BibitemShut {NoStop}%
\end{thebibliography}%



\end{document}